\def\REVIEW{0}%

\documentclass[preprint,journal]{vgtc}            

\onlineid{1382}

\vgtccategory{Research}

\title{Fact-Check Your Information (FYI): A Design Probe to Understand How People Actually Fact-Check Data-Driven Articles}

\author{%
  \authororcid{Nguyen-Truong Thinh$^*$}{0009-0006-5441-7057},
  \authororcid{Yuxuan Du$^*$}{0009-0000-1299-7520},
  \authororcid{Phongsakon Mark Konrad}{0009-0004-2521-7879},
  \authororcid{Arpit Narechania}{0000-0001-6980-3686}
}

\authorfooter{
\item
    Nguyen-Truong is with Holistics Software and is an incoming M.Phil. student at The Hong Kong University of Science and Technology.
  	E-mail: truongthinh.nguyen03@gmail.com
\item
  	Yuxuan is an Independent Contributor.
  	E-mail: yuxuan.du.sherry@gmail.com
\item
    Mark is with the University of Southern Denmark.
    E-mail: phkon23@student.sdu.dk
\item
    Arpit is with The Hong Kong University of Science and Technology.
  	E-mail: arpit@ust.hk    
\item[] $^*$ $=$ Authors contributed equally to this work.}

\abstract{%
Data-driven journalism and policy reports frequently rely on statements grounded in statistical evidence, referred to as \textit{data claims}.
Verifying such a claim requires connecting it to the underlying structured dataset.
However, existing systems typically isolate automated fact-checking from manual data exploration, leaving it unclear how readers coordinate AI assistance with manual inspection of the evidence in practice.
We present \app, a browser extension that embeds fact-checking in the reading environment, and use it as a design probe to study how people detect, verify, and determine the validity of data claims against the underlying dataset. \app provides four complementary tools spanning the spectrum from full automation to manual data exploration.
In an exploratory study ($N = 22$), participants used FYI to fact-check claims in a data-driven article.
We find that participants adopted three distinct workflow archetypes---AI-first with manual confirmation, manual-first with AI supplement, and parallel co-review---with visualization serving as the primary mechanism for auditing AI conclusions.
Trust in AI shifted dynamically, growing when multiple tools converged and eroding when AI outputs were inconsistent.
These findings suggest that fact-checking systems should treat AI as a starting point that human verification complements rather than a definitive authority, elevate visualization as a core verification capability, and support flexible, user-driven workflows.
We release \app as open-source software for further research at \url{https://github.com/DataVisards/FYI}.
}

\keywords{Data claims, fact-checking, human-AI interaction, trust calibration, visualization, large language models, design probe.}

\teaser{
  \centering
  \includegraphics[width=\linewidth, alt={A browser window with a news article on the left and the FYI side panel filling the rest. In the article, the sentence claiming Drama has the highest average IMDb rating is highlighted, with a Mark as claim button beside it and a Detect claims with AI button below. In the panel, AutoCheck returns MISLEADING at moderate confidence and sets the claimed Drama is number 1 against the actual Thriller at 6.9, above a bar chart of average rating by genre in which Drama is the shortest of the five bars. Below it an AI Chat exchange repeats the correction, a Table Explorer lists movie titles with gross, budget, year and rating, and a Chart Builder is configured with Genre on X and average IMDb rating on Y. On the right a verdict form has Misleading selected, confidence 6 of 7, severity 5 of 7, and a rewritten version of the claim naming Thriller instead of Drama.}]{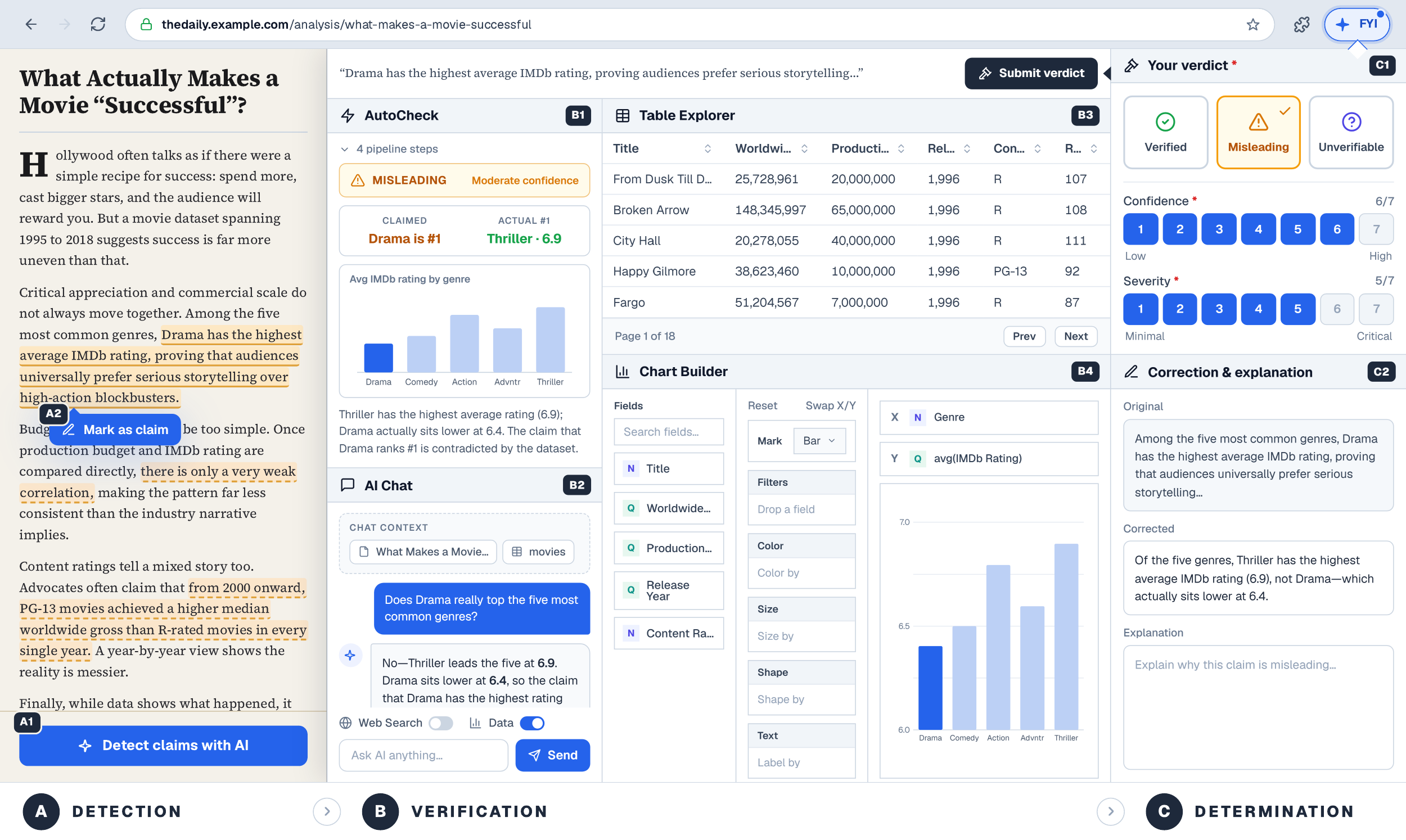}
  \caption{The interface of \app, a browser extension that enables in-situ data claim fact-checking directly alongside the article under review. The unified side panel supports a complete mixed-initiative pipeline: \textbf{(A) Detection} via (A1) AI detection or (A2) manual identification; \textbf{(B) Verification} through four complementary modalities: (B1) \toolAutoCheck, (B2) \toolAIChat, (B3) \toolTable, and (B4) \toolChart; and \textbf{(C) Determination} via (C1) user-authored verdicts and (C2) corrections and explanations.}
  \label{fig:teaser}
}

\PassOptionsToPackage{svgnames}{xcolor}

\graphicspath{{figs/}{figures/}{pictures/}{images/}{./}}

\AtEndPreamble{\hypersetup{pdflang=en-US}}

\AtEndPreamble{%
  \captionsetup{labelsep=period}%
}

\usepackage{tabu}
\usepackage{booktabs}
\usepackage{multirow}
\usepackage{makecell}
\usepackage{colortbl}
\usepackage{rotating}

\usepackage{mathptmx}                  %
\usepackage{xcolor}
\usepackage{soul}
\usepackage{enumitem}
\usepackage{quoting}
\usepackage{listings}
\usepackage{fontawesome5}
\usepackage{setspace}
\usepackage{amssymb}

\usepackage{lipsum}                    %
\usepackage{mwe}                       %

\usepackage{xcolor}
\usepackage{ulem}
\providecommand{\REVIEW}{1}
\newif\ifreview
\ifnum\REVIEW=1\relax \reviewtrue \else \reviewfalse \fi
\ifreview
  \NewDocumentCommand{\added}{o m}{\textcolor{blue}{#2}}
  \NewDocumentCommand{\deleted}{o m}{\textcolor{red}{\sout{#2}}}
  \NewDocumentCommand{\replaced}{o m m}{\textcolor{blue}{#2}\textcolor{red}{\sout{#3}}}
  \NewDocumentCommand{\highlight}{o m}{\colorbox{yellow}{#2}}
\else
  \NewDocumentCommand{\added}{o m}{#2}
  \NewDocumentCommand{\deleted}{o m}{}
  \NewDocumentCommand{\replaced}{o m m}{#2}
  \NewDocumentCommand{\highlight}{o m}{#2}
\fi

\usepackage[switch]{lineno}

\AtEndPreamble{%
  \crefname{figure}{Fig.}{Figs.}
  \Crefname{figure}{Figure}{Figures}
  \crefname{table}{Tab.}{Tabs.}
  \Crefname{table}{Table}{Tables}
  \crefname{section}{Sec.}{Secs.}
  \crefname{equation}{Eq.}{Eqs.}
}

\newcommand{\app}{FYI\xspace}

\newcommand{\bpstart}[1]{\vspace{0.15cm}\noindent\textbf{#1}}

\definecolor{colorAutoCheck}{HTML}{0072B2}  %
\definecolor{colorAIChat}{HTML}{CC79A7}     %
\definecolor{colorTable}{HTML}{009E73}      %
\definecolor{colorChart}{HTML}{E69F00}      %

\newcommand{\toolpill}[3]{%
  \mbox{%
    \setlength{\fboxsep}{1.2pt}%
    \smash{%
      \colorbox{#1}{%
        \rule{0pt}{0.7em}%
        \kern1pt
        \raisebox{-1pt}{\includegraphics[height=0.7em]{icons/#2.pdf}}%
        \,\textcolor{white}{\small\textsf{#3}}%
        \kern1pt
      }%
    }%
    \vphantom{Ag}%
  }%
}

\newcommand{\toolAutoCheck}{\toolpill{colorAutoCheck}{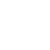}{Auto Check}\xspace}
\newcommand{\toolAIChat}{\toolpill{colorAIChat}{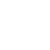}{AI Chat}\xspace}
\newcommand{\toolTable}{\toolpill{colorTable}{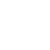}{Table Explorer}\xspace}
\newcommand{\toolChart}{\toolpill{colorChart}{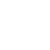}{Chart Builder}\xspace}

\newlength{\toolpillW}
\newcommand{\toolpillfix}[3]{\toolpill{#1}{#2}{\makebox[\toolpillW][l]{#3}}}
\newcommand{\toolAutoCheckT}{\toolpillfix{colorAutoCheck}{zap}{Auto Check}}
\newcommand{\toolAIChatT}{\toolpillfix{colorAIChat}{bot}{AI Chat}}
\newcommand{\toolTableT}{\toolpillfix{colorTable}{table2}{Table Explorer}}
\newcommand{\toolChartT}{\toolpillfix{colorChart}{chart-area}{Chart Builder}}

\newcommand{\cut}[1]{}

\AtBeginDocument{\hbadness=10000\vbadness=10000\hfuzz=100pt\vfuzz=100pt}

\begin{document}
\raggedbottom
\addtolength{\textheight}{7pt}
\raggedbottom
\addtolength{\textheight}{7pt}

\firstsection{Introduction}
\maketitle

Consider a reader encountering the following statistic in a public health report: \textit{``The risk of death involving COVID-19 was consistently lower for people who had received two vaccinations compared to one or no vaccination''}~\cite{ons2021coviddeaths}.
This is a data claim, a statement grounded in quantitative evidence from structured datasets~\cite{fu2024data}, routinely found in data journalism and policy reporting~\cite{zamith2019transparency}. 
Unlike textual fact-checking, which relies on finding corroborating sources, verifying a data claim requires specialized tools to aggregate, compare, and interpret structured data.
This illusion of rigor can make misrepresented statistics appear credible, especially when readers lack the time or dedicated tools to inspect the underlying data.

Systematically addressing this risk requires supporting readers through the established fact-checking pipeline: claim \textit{detection} (identifying check-worthy statements), \textit{verification} (evaluating evidence against the dataset), and \textit{determination} (synthesizing that evidence into a judgment)~\cite{guo2022survey,graves2018understanding}.
Each stage combines rigorous data processing with subjective reasoning, so navigating the pipeline demands both computational assistance and manual inspection of the data.

Existing systems span the human-AI spectrum but remain fragmented. 
Fully automated pipelines are unreliable on precise numerical claims~\cite{pesaranghader2026hallucination,yuan2023large}, interactive analytics platforms impose a high cognitive burden~\cite{kim2024datadive}, crowdsourced approaches lack analytical depth~\cite{lloyd2025beyond}, and lightweight browser extensions provide only shallow credibility signals~\cite{jahanbakhsh2024browser,botnevik2020brenda} (see \cref{sec:related-work} for a detailed review).
No unified environment lets users flexibly move between AI-generated outputs and direct data examination across the full fact-checking pipeline.

Mixed-initiative systems introduce a fundamental tension. As agency shifts towards AI, users may accept outputs without examining the data (automation bias~\cite{goddard2012automation,lee2025impact,liu2026behavioral}), while full human control risks cognitive overload~\cite{lane2014effect,boldova2024cognitive}.
Users must actively decide how much agency to delegate at each step, constituting a form of trust calibration that can shift across phases and claims, and in response to prior AI reliability.
How users calibrate this trust when both modalities are available remains poorly understood in data claim fact-checking.

These tensions point to an empirical gap in understanding how people actually behave when fact-checking data claims with AI-assisted and manual modalities simultaneously available.
Studying these behaviors requires an integrated and instrumented artifact that makes cross-modal workflows observable. 

\bpstart{}To address this gap, we pursue the following research questions:
\begin{itemize}[nosep]
    \item \textbf{RQ1} (Detection): How do users detect check-worthy data claims?
    \item \textbf{RQ2} (Verification): How do users verify data claims?
    \item \textbf{RQ3} (Determination): How do users reach and communicate their fact-checking verdict?
    \item \textbf{RQ4} (Human-AI): How do users balance trust in AI against staying in control throughout the fact-checking process?
\end{itemize}

\bpstart{}To answer these questions, we designed Fact-check Your Information (FYI), a browser extension that embeds the fact-checking workflow directly within the reading environment.
FYI is scoped to dataset-grounded data-claim fact-checking for data-driven articles, whose authors are expected to disclose the underlying dataset as a downloadable file or linked source. It is intended for laypersons with basic data and visualization literacy who consume such articles online and can inspect tables or interpret charts, but who do not necessarily have access to professional fact-checking infrastructure. FYI supports the detection--verification--determination workflow: users can manually highlight claims or use AI detection; verify claims through a fully automated pipeline ({\toolAutoCheck}), a conversational AI agent ({\toolAIChat}), an interactive table ({\toolTable}), and a visualization builder ({\toolChart}); and submit verdicts with corrections where appropriate.

Using FYI as a design probe, we conducted an exploratory user study (N=22) in which participants freely combined these tools to fact-check data claims in a realistic data-driven article.
By logging fine-grained interaction sequences, we turn the reading session itself into an observational window into users’ fact-checking process. 
Our primary contribution is an exploratory design-probe study of mixed-initiative data-claim fact-checking. Specifically, we contribute:
\begin{enumerate}[nosep]
    \item Empirical characterization of how laypersons compose fact-checking strategies, move across AI-assisted and manual tools, and calibrate trust across modalities.
    \item Design implications for future mixed-initiative fact-checking systems.
    \item FYI, an open-source prototype contributed to the community as a testbed for future data fact-checking research (\url{https://github.com/DataVisards/FYI}).
\end{enumerate}

\section{Related Work}
\label{sec:related-work}

\definecolor{cYes}{HTML}{C6EFCE}
\definecolor{cNo}{HTML}{FCE4EC}
\definecolor{cPartial}{HTML}{FFF3CD}
\definecolor{catA}{HTML}{E8E8F0}   %
\definecolor{catB}{HTML}{E0EEEA}   %
\definecolor{catC}{HTML}{F2E8DE}   %
\definecolor{catD}{HTML}{E2E9F3}   %

\DeclareRobustCommand{\dimpill}[2]{%
  \mbox{%
    \setlength{\fboxsep}{1.2pt}%
    \smash{\colorbox{#1}{\kern1pt\small\vphantom{Ag}#2\kern1pt}}%
    \vphantom{Ag}%
  }%
}

\newcommand{\cmark}{\cellcolor{cYes}\textbf{Y}}
\newcommand{\xmark}{\cellcolor{cNo}\textbf{N}}
\newcommand{\pmark}{\cellcolor{cPartial}\textbf{P}}

\begin{table*}[t!]
  \centering
  \caption{Comparison of data fact-checking systems across four dimensions:
    \dimpill{catA}{Claim Detection}~(M1),
    \dimpill{catB}{Claim Verification}~(M2),
    \dimpill{catC}{Claim Determination}~(M3), and
    \dimpill{catD}{Human Agency}~(M4). The table compares systems relevant to data-claim fact-checking and adjacent evidence-checking tasks. Columns indicate which parts of the detection--verification--determination workflow each system supports, without implying that all systems are end-to-end fact-checking systems. Systems are grouped by interaction model---fully-automated NLP systems (top) and human-in-the-loop systems (bottom)---to reflect design intent instead of comparing their capability.}
  \label{tab:comparison}

  \smallskip
  {\small \textbf{Y}\,=\,Supported \quad
         \textbf{P}\,=\,Partially Supported \quad
         \textbf{N}\,=\,Not Supported}

  \vspace{6pt}
  \footnotesize
  \setlength{\tabcolsep}{3pt}
  \renewcommand{\arraystretch}{1.15}
  \resizebox{\textwidth}{!}{%
  \setlength{\aboverulesep}{0pt}
  \setlength{\belowrulesep}{0pt}
  \setlength{\extrarowheight}{2pt}

  \begin{tabular}{@{}l l
      >{\columncolor{catA}}c >{\columncolor{catA}}c
      >{\columncolor{catB}}c >{\columncolor{catB}}c >{\columncolor{catB}}c >{\columncolor{catB}}c
      >{\columncolor{catC}}c >{\columncolor{catC}}c
      >{\columncolor{catD}}c >{\columncolor{catD}}c}
    \toprule
    & &
    \multicolumn{2}{>{\columncolor{catA}}c}{\textbf{Claim Detection}} &
    \multicolumn{4}{>{\columncolor{catB}}c}{\textbf{Claim Verification}} &
    \multicolumn{2}{>{\columncolor{catC}}c}{\textbf{Claim Determination}} &
    \multicolumn{2}{>{\columncolor{catD}}c}{\textbf{Human Agency}} \\
    & &
    \multicolumn{2}{>{\columncolor{catA}}c}{\scriptsize\textit{M1}} &
    \multicolumn{4}{>{\columncolor{catB}}c}{\scriptsize\textit{M2}} &
    \multicolumn{2}{>{\columncolor{catC}}c}{\scriptsize\textit{M3}} &
    \multicolumn{2}{>{\columncolor{catD}}c}{\scriptsize\textit{M4}} \\
    \cmidrule(lr){3-4} \cmidrule(lr){5-8} \cmidrule(lr){9-10} \cmidrule(lr){11-12}
    \textbf{System} & \textbf{Venue} &
    \rotatebox{70}{\makecell[l]{Automated\\Detection}} &
    \rotatebox{70}{\makecell[l]{Manual\\Selection}} &
    \rotatebox{70}{\makecell[l]{Automated\\Verification}} &
    \rotatebox{70}{\makecell[l]{NL\\Dialog}} &
    \rotatebox{70}{\makecell[l]{Tabular\\Inspection}} &
    \rotatebox{70}{\makecell[l]{Visual\\Analytics}} &
    \rotatebox{70}{\makecell[l]{Human\\Verdict}} &
    \rotatebox{70}{\makecell[l]{In-situ\\Annotation}} &
    \rotatebox{70}{\makecell[l]{AI\\Override}} &
    \rotatebox{70}{\makecell[l]{Multi-Tool\\Orchestration}} \\
    \midrule
    \multicolumn{12}{@{}l}{\textbf{Fully-Automated NLP Systems}}\\
    ClaimBuster \cite{hassan2017claimbuster}
      & KDD\,'17
      & \cmark & \pmark & \cmark & \xmark & \xmark & \xmark & \pmark & \xmark & \xmark & \xmark \\
    TabFact \cite{chen2020tabfact}
      & ICLR\,'20
      & \xmark & \xmark & \cmark & \xmark & \cmark & \xmark & \xmark & \xmark & \xmark & \xmark \\
    PASTA \cite{gu2022pasta}
      & EMNLP\,'22
      & \xmark & \xmark & \cmark & \xmark & \cmark & \xmark & \xmark & \xmark & \xmark & \xmark \\
    DATER \cite{ye2023dater}
      & SIGIR\,'23
      & \xmark & \xmark & \cmark & \xmark & \cmark & \xmark & \xmark & \xmark & \xmark & \xmark \\
    Binder \cite{cheng2023binder}
      & ICLR\,'23
      & \xmark & \xmark & \cmark & \xmark & \cmark & \xmark & \xmark & \xmark & \pmark & \xmark \\
    Chain-of-Table \cite{wang2024chainoftable}
      & ICLR\,'24
      & \xmark & \xmark & \cmark & \xmark & \cmark & \xmark & \xmark & \xmark & \xmark & \xmark \\
    RePanda \cite{chegini2025repanda}
      & ACL\,'25
      & \xmark & \xmark & \cmark & \xmark & \cmark & \xmark & \xmark & \xmark & \xmark & \xmark \\
    \midrule
    \multicolumn{12}{@{}l}{\textbf{Human-in-the-Loop Systems}}\\
    Believe it or not \cite{nguyen2018believe}
      & UIST\,'18
      & \xmark & \cmark & \cmark & \xmark & \xmark & \xmark & \cmark & \xmark & \cmark & \xmark \\
    Scrutinizer \cite{karagiannis2020scrutinizer}
      & VLDB\,'20
      & \cmark & \pmark & \pmark & \pmark & \cmark & \xmark & \cmark & \xmark & \cmark & \xmark \\
    StatCheck \cite{balalau2022statcheck}
      & CIKM\,'22
      & \cmark & \pmark & \cmark & \xmark & \cmark & \xmark & \xmark & \xmark & \pmark & \xmark \\
    CrossData \cite{chen2022crossdata}
      & CHI\,'22
      & \cmark & \cmark & \pmark & \xmark & \cmark & \cmark & \xmark & \cmark & \cmark & \cmark \\
    DataTales \cite{sultanum2023datatales}
      & VIS\,'23
      & \xmark & \cmark & \pmark & \xmark & \pmark & \cmark & \xmark & \xmark & \cmark & \cmark \\
    DataDive \cite{kim2024datadive}
      & IUI\,'24
      & \xmark & \cmark & \pmark & \pmark & \cmark & \cmark & \xmark & \xmark & \cmark & \cmark \\
    EmphasisChecker \cite{choi2024emphasischecker}
      & TVCG\,'24
      & \cmark & \cmark & \cmark & \xmark & \cmark & \cmark & \xmark & \cmark & \cmark & \cmark \\
    Aletheia \cite{fu2024data}
      & UIST\,'24
      & \cmark & \cmark & \cmark & \xmark & \cmark & \cmark & \cmark & \cmark & \pmark & \cmark \\
    MisVisFix \cite{das2025misvisfix}
      & TVCG\,'25
      & \cmark & \cmark & \cmark & \cmark & \xmark & \cmark & \pmark & \cmark & \cmark & \cmark \\
    T-REX \cite{horstmann2025trex}
      & ECML\,'25
      & \xmark & \cmark & \cmark & \xmark & \cmark & \cmark & \xmark & \cmark & \xmark & \cmark \\
    \midrule
    \textbf{\textit{FYI (ours)}}
      & ---
      & \cmark & \cmark & \cmark & \cmark & \cmark & \cmark & \cmark & \cmark & \cmark & \cmark \\
    \bottomrule
  \end{tabular}
  }%
\end{table*}

\subsection{From Text Claims to Data Claims}

The foundation of automated fact-checking was established primarily within the NLP community, focusing almost exclusively on unstructured text. 
The standard process is organized into a three-stage pipeline~\cite{guo2022survey}: \textit{detection} of check-worthy statements, \textit{verification} against retrieved textual evidence, and \textit{determination} of a final veracity label.
Early benchmarks such as FEVER~\cite{thorne2018fever}, LIAR~\cite{wang2017liar}, and SciFact~\cite{wadden2020scifact} operated strictly within this framework.
For text claims, detection typically involves identifying rhetorical markers or subjective versus objective framing to determine whether a statement is worthy of investigation~\cite{hyland2005stance}.
Subsequently, the verification phase is treated fundamentally as a natural language inference (NLI) task, where the core challenge is semantic matching to determine whether a retrieved piece of text entails or refutes the detected claim~\cite{hanselowski2018neural}.

However, \textit{data claims}, statements grounded in quantitative evidence drawn from structured datasets, require fundamentally different approaches.
Detecting a data claim requires identifying numerical assertions, statistical summaries, or comparative trends embedded within prose, and recognizing that these statements implicitly refer to an underlying dataset~\cite{vlachos2015identification}. 
Once a claim is detected, verifying a data claim demands a combination of linguistic interpretation and precise analytical operations such as comparing, counting, and aggregating over table rows or visual marks~\cite{chen2020tabfact} to derive the answer from the raw data. 
Thus, fact-checking of data claims has inherently higher computational and cognitive complexity than pure text processing.

\subsection{Paradigms of Fact-Checking Systems}

In this section, we review systems that support data-claim fact-checking or closely related evidence-checking tasks involving structured data, tables, or visualizations. To address the complexity of data claims, existing system designs have largely diverged into paradigms along the human-AI spectrum.

One is the fully automated paradigm, which aims to minimize user effort by delegating complex reasoning entirely to AI.
For the detection phase, models like ClaimBuster~\cite{hassan2017claimbuster} and subsequent LLM-based extractors~\cite{metropolitansky2025towards} attempt to flag check-worthy statements automatically without user intervention.
For verification, the field has shifted toward neural-symbolic decomposition strategies. Binder~\cite{cheng2023binder} parses claims into executable SQL or Python expressions, while DATER~\cite{ye2023dater} decomposes tables into focused sub-tables for LLM reasoning, and Chain-of-Table~\cite{wang2024chainoftable} applies iterative table operations to evolve evidence through a reasoning chain.
Recent end-to-end architectures, such as Aletheia~\cite{fu2024data}, attempt to automate the entire lifecycle from semantic parsing to the generation of interactive data evidence representations, and multi-agent systems like Thucy~\cite{theologitis2026thucy} deploy specialized LLM agents to verify claims across relational databases with executable SQL evidence (achieving 94.3\% on TabFact).
Despite their computational power, these fully automated systems primarily operate without a human-in-the-loop mechanism. When automated reasoning produces an incorrect or hallucinated result, users are unable to contest the output~\cite{buholayka2024reference}. Although Aletheia provides interactive widgets for overriding AI-inferred filters, its user agency remains limited to error correction rather than open-ended exploration.

In contrast, the interactive paradigm prioritizes direct data inspection. Some systems target verification problems adjacent to article-level fact-checking and demonstrate the role of visual evidence in checking data-related statements. For example,
DataTales~\cite{sultanum2023datatales} supports LLM-assisted authoring of data-driven articles with integrated visual evidence, while EmphasisChecker~\cite{choi2024emphasischecker} helps users construct analytical charts to check text-chart consistency. They both show how visual evidence can support inspection of data-related claims, even though their primary tasks differ.
Meanwhile, visual overviews and raw tabular inspection serve complementary roles. Kim et al. ~\cite{kim2018texttable} suggested that rigorous verification often requires users to drill down into specific table cells to resolve granular ambiguities.
However, this paradigm assumes that the user has already detected the claim and formulated a verification strategy. 
Further, relying purely on interactive tools to navigate raw data and construct visual evidence imposes relatively high analytical burden and expertise requirements~\cite{boldova2024cognitive}.

Recently, mixed-initiative systems have sought to bridge the gap. At the organizational scale, Scrutinizer~\cite{karagiannis2020scrutinizer} demonstrated a structured mixed-initiative approach in which the system proposes SQL query fragments that domain experts validate or correct, with classifiers improving through active learning from accumulated human feedback, reducing verification time by over 50\%.
For individual users, StatCheck~\cite{balalau2022statcheck} prioritizes user agency by retrieving relevant statistical databases for journalists, leaving the final verification to the human.
Similarly, real-time streaming interfaces like T-REX~\cite{horstmann2025trex} highlight relevant table cells alongside LLM reasoning to make automated verification more transparent.
Despite these advancements, the current systems for data claim fact-checking remain highly fragmented. Specifically, they are either fully automated, which strips user control, or highly manual, which overwhelms cognitive capacity. We currently lack unified environments that support a flexible workflow, where users can seamlessly transition between automated and human-in-the-loop fact-checking.

\subsection{Human-AI Sensemaking and Trust Calibration}

Bridging the gap between fully automated pipelines and high-effort manual analytics requires framing data fact-checking as a complex sensemaking process. 
Foundational models of sensemaking describe data analysis as an iterative cycle of information search, hypothesis establishment, and evidence synthesis~\cite{pirolli2005sensemaking}. 
In the data-claim fact-checking context, this loop spans both the detection and verification phases. It involves mapping claims to structured evidence and iteratively inspecting tables or visualizations to test interpretations.

However, introducing AI into the fact-checking process fundamentally alters user agency. 
A key challenge in AI-assisted analysis is that users over-rely on automated outputs instead of carefully seeking and checking information themselves~\cite{goddard2012automation}. 
In fact-checking scenarios, users might passively accept the AI verdict or an auto-generated chart without verifying its provenance, even when the underlying reasoning is flawed~\cite{chae2024perceiving}. 
Conversely, if an AI fails to detect a claim or cannot explain its verification logic through transparent evidence, users may lose confidence and exhibit algorithmic aversion~\cite{jonesjang2023ai}, entirely rejecting valid automated assistance and reverting to a fully manual check.

Therefore, effective mixed-initiative systems must be designed to support trust calibration, the process by which users align their reliance on the AI with the system's actual reliability~\cite{yin2019understanding}. 
This requires interactive override capabilities that allow users to recover from AI errors and re-exert agency~\cite{nguyen2018believe}. 
Critically, the level of reliance may differ across the detection and verification phases~\cite{zhang2020adaptive}. For instance, AI claim detection may be accurate yet incomplete, while AI verdicts may be fluent but numerically incorrect. Users, therefore, need to calibrate trust based on the detailed scenarios.
Furthermore, the benefits of explainability in automated fact-checking remain contested; Lim and Perrault~\cite{lim2023xai} found that only one out of five tested XAI modalities marginally improved user performance, while graphical explanations may paradoxically anchor users to the AI's framing rather than encouraging independent verification.
A recent review of human-AI decision support~\cite{samu2026teammate} further warns of a ``fluency trap'' where conversational AI interfaces inflate perceived understanding and trust without reliably improving decision quality, underscoring the need to evaluate whether multi-modal verification tools genuinely improve judgment or merely shift reliance patterns.

Although existing research has studied the fact-checking workflows and tool use of professionals~\cite{juneja2022human}, empirical studies on laypeople remain exceedingly scarce. 
Specifically, because users rarely have access to a fully integrated toolset, there is a profound deficit in our understanding of how they actually behave. 
We lack empirical evidence on how users orchestrate strategies, divide agency, and calibrate trust when automated and interactive tools are simultaneously available. 
This critical empirical gap motivates the deployment of FYI as an in-situ design probe to observe and characterize these multi-modal behavioral dynamics and sensemaking strategies in real-world reading environments.

\subsection{Design Gaps and Positioning}
\label{sec:rw-matrix}

As summarized in \cref{tab:comparison}, the comparison is structured along the detection--verification--determination pipeline extended with a human agency dimension, which is central to mixed-initiative fact-checking. 

The upper part of the table shows that NLP systems typically provide limited support for human agency and determination except for Binder's partial AI override or ClaimBuster's limited manual inputs.  
Furthermore, they mostly treat claim detection and verification as an automated preprocessing step rather than a user-driven behavior. 
As a result, interactive operations, user-authored verdicts, and multi-tool orchestration often fall outside their design scope.

The lower part of the table shows a complementary pattern.  Human-in-the-loop systems provide stronger support for user agency. For example,
CrossData and EmphasisChecker successfully empower users with multi-tool orchestration, AI overrides, and in-situ annotation, while Aletheia provides multi-tool evidence and partial AI override through interactive filter correction.
However, this agency often shifts more analytical work to users. For instance, DataTales and DataDive offer only partial support for AI-assisted fact-checking, so users still need to formulate analyses and inspect outputs. 
More critically, across both the NLP and human-in-the-loop paradigms, natural language dialog remains largely absent. While Scrutinizer offers structured query-based interaction and DataDive allows optional free-form text input, only MisVisFix~\cite{das2025misvisfix} provides a multi-turn chat interface.

This comparison motivates FYI as an integrated design-probe prototype for our study. 
By combining natural language dialog ({\toolAIChat}), automated verification ({\toolAutoCheck}), tabular inspection ({\toolTable}), and visual analytics ({\toolChart}), FYI brings together modalities that are typically separated across prior systems, making it possible to empirically observe how users orchestrate these previously fragmented modalities to fact-check data claims.

\section{FYI: A Design Probe}
\label{sec:system-design}

\subsection{Design Goals}
\label{sec:design-goals}

To observe how readers fact-check data claims when automated and manual tools are simultaneously available, FYI needed to keep the workflow within the reading context, expose multiple available verification modalities, and preserve user agency while recording the resulting process. Drawing on prior work in automated fact-checking~\cite{guo2022survey,fu2024data}, interactive data exploration~\cite{kim2024datadive,huang2026facilitating}, and analytical provenance~\cite{narechania2024provenancewidgets,block2022influence}, we translate these requirements into three design goals.

\bpstart{DG\textsubscript{1}: In-situ integration.}
Existing fact-checking tools often require users to leave their reading environment~\cite{schultz2012truth}, incurring context-switching costs~\cite{rubinstein2001executive}.
\app should embed the full investigation workspace in the browser as a side panel, allowing users to verify claims while reading the article.

\bpstart{DG\textsubscript{2}: Multi-modal verification.}
Data claim verification is rarely reducible to a single technique~\cite{fu2024data}: AI pipelines accelerate evidence retrieval but may hallucinate~\cite{pesaranghader2026hallucination,yuan2023large}, tabular inspection reveals precise values, visualizations expose distributional patterns, and web search provides external corroboration~\cite{aslett2024online}.
\app should offer multiple complementary tools so that users can triangulate evidence across modalities.

\bpstart{DG\textsubscript{3}: Human agency and provenance.}
Recent work highlights the risk of over-reliance on AI outputs during fact-checking~\cite{lee2025impact,liu2026behavioral,deverna2024llmfactcheckbehavior}.
\app should keep the user as the primary decision-maker: AI provides evidence and suggestions, but verdicts are always user-submitted.
The system should also log all interactions to enable post-hoc analysis of verification workflows and trust calibration.

Together, these goals shape \app as a design probe that supports in-situ fact-checking, multi-modal verification, and user-authored determination while making users’ interaction behaviors observable.

\subsection{Prototype Overview}
\label{sec:system-overview}

\app is a browser extension that hosts its fact-checking workspace in a side panel beside the article under review. Targeting data-driven articles, FYI assumes the underlying dataset is available, either as a downloadable file or a linked source---a transparency norm in data journalism~\cite{zamith2019transparency}. The reader uploads this dataset as grounding data, shared between the AI's analysis and the reader's own inspection. The reader can then confirm each AI suggestion against this data, viewing it as tables and charts, instead of accepting it on the AI's word.

\bpstart{}The user workflow consists of three stages:
\begin{enumerate}[nosep,leftmargin=*]
  \item \textbf{Detection.} The user activates \app on an article page, uploads CSV datasets as grounding data, and initiates claim detection either via the AI pipeline or by manually highlighting passages. Detected claims are highlighted in the article (\cref{sec:claim-detection}).
  \item \textbf{Verification.} The user selects a claim and investigates it using four complementary tools (\cref{sec:tools}), triangulating evidence across AI and manual modalities.
  \item \textbf{Determination.} The user submits a judgment (verified, misleading, or unverifiable) with confidence and severity ratings, and optionally provides a textual correction for misleading claims (\cref{sec:verdict}).
\end{enumerate}

\bpstart{}All AI-powered components (claim detection, \toolAutoCheck, and \toolAIChat) use GPT-4.1~\cite{openai2025gpt41}, except web search, which uses Perplexity Sonar; all prompts are included in the supplemental materials.

\subsection{Claim Detection}
\label{sec:claim-detection}

Claim detection supports both automated and manual pathways, spanning the human-AI spectrum.
AI-powered detection offers speed and coverage, while manual curation preserves user agency and captures claims that automated methods may miss---such as implicit comparisons or domain-specific assertions.
Providing both enables observation of how users divide detection labor between themselves and AI.

\bpstart{AI-powered detection.}
Data claims are often embedded in flowing prose in ways that readers may not immediately recognize as verifiable~\cite{fu2024data}, so automated detection lowers the entry barrier and supports broad coverage.
An LLM classifies whether each sentence is a potential data claim, which is defined as a natural-language statement whose veracity depends on a specific dataset.
Detected claims are highlighted directly in the article and listed in the side panel for investigation.

\bpstart{Manual claim detection.}
AI detection can miss claims that depend on context, domain knowledge, or subjective judgment about what is worth checking, so users need to supplement and override.
Users can select any text passage on the article page, triggering a floating toolbar with a ``Mark as claim'' option (\textbf{DG\textsubscript{3}}).
User-added claims always take priority: when a manual claim overlaps an AI-detected claim in the same text region, the AI claim is automatically suppressed.
Users can also dismiss AI-detected claims they consider irrelevant, but AI cannot modify or remove user-added claims.
This asymmetry ensures that human judgment always supersedes automated detection, while dismissal patterns provide a behavioral signal of disagreement with AI.

\subsection{Verification Tools}
\label{sec:tools}

The investigation workspace provides four complementary tools (\textbf{DG\textsubscript{2}}), each accessible as a tab within the side panel.
Users can switch between tools freely, and the system tracks tool-usage sequences and dwell time per claim (\textbf{DG\textsubscript{3}}).

The four tools span a spectrum from fully automated (\toolAutoCheck) through user-directed AI (\toolAIChat) to fully manual (\toolTable, \toolChart), enabling observation of how users compose verification strategies.

\subsubsection{Auto Check}
\label{sec:auto-check}
\toolAutoCheck provides a zero-effort baseline, executing a four-step streaming pipeline: (1)~resolving references to make the claim self-contained, (2)~generating and executing verification code (Binder-inspired~\cite{cheng2023binder}), (3)~producing an interactive Vega-Lite evidence chart, and (4)~outputting a verdict badge with confidence, a claimed-vs-actual comparison, and natural-language reasoning.
Steps are available in a collapsible view, and a ``Follow up in AI Chat'' button seeds deeper exploration in \toolAIChat.

\subsubsection{AI Chat}
\label{sec:ask-ai}
Because \toolAutoCheck produces a single predetermined analysis, users often need to ask follow-up questions, explore alternative interpretations, or seek external corroboration---tasks that require flexible, user-directed AI assistance.
\toolAIChat provides a multi-turn conversational interface for this open-ended verification.
The system prompt includes the article content, the selected claim, and metadata for any uploaded datasets.
Users can toggle two capabilities to control the scope of AI involvement:

\begin{itemize}[nosep,leftmargin=*]
  \item \textbf{Web Search} enables the model to query external sources for corroboration beyond the dataset; results are rendered as numbered inline citations with clickable source links.
  \item \textbf{Data Analysis} grants the model access to Python code execution via client-side Pyodide, with datasets pre-loaded as pandas DataFrames. This allows multi-step statistical analysis, aggregation, and computation within the conversation.
\end{itemize}

\subsubsection{Table Explorer}
\label{sec:inspect-table}
AI-generated summaries can obscure the underlying evidence, so \toolTable provides direct, unmediated access to the raw dataset, enabling users to inspect specific values rather than relying on AI interpretations.
Users can sort by any column, apply categorical filters (checkbox multi-select) or quantitative filters (range sliders), and paginate through large datasets.

\subsubsection{Chart Builder}
\label{sec:create-vis}
Both \toolTable and \toolChart operate without AI, but \toolChart demands the highest analytical effort: users must decide what to plot, how to encode it, and how to interpret the result.
Following the shelf-based encoding paradigm of Voyager~\cite{wongsuphasawat2016voyager,wongsuphasawat2017voyager2}, users build Vega-Lite charts by dragging dataset fields onto encoding shelves (x, y, color, size), selecting mark types (bar, line, point, area), and applying aggregation functions (e.g., average, median, count) and filters.
Multiple chart tabs allow parallel exploration of different hypotheses for the same claim.

\subsection{Verdict and Correction}
\label{sec:verdict}

The verdict form serves as both a decision endpoint and a data collection instrument. By capturing not only the judgment but also the reasoning and tool attribution, it enables post-hoc analysis of how different evidence sources map to different decision outcomes.
Users submit a judgment (\emph{verified}, \emph{misleading}, or \emph{unverifiable}), a 7-point confidence rating, and a ranked list of which tools were helpful during the investigation (ordered from most to least helpful).
For misleading and unverifiable claims, users additionally provide a severity rating (1--7) and an explanation; misleading claims also support an inline correction where users edit the original claim text to reflect the accurate version.

\subsection{Interaction Logging}
\label{sec:interaction-logging}

To enable the design probe analysis (\textbf{DG\textsubscript{3}}), \app records every user action as a timestamped event.
The event schema covers 25 event types spanning the full pipeline: claim detection and dismissal, tool invocations and toggle changes, data exploration actions (sorting, filtering, chart encoding changes), chat exchanges (messages, responses, tool calls), and verdict submissions with all associated metadata.
At session completion, the system exports a JSON artifact containing the full event log, all claims and verdicts, and session summary metrics.
This granular provenance data enables reconstruction of each participant's complete verification workflow---which tools were used for which claims, in what order, and with what outcomes---supporting the behavioral analyses reported in \cref{sec:results}.

\section{Study}
\label{sec:study}

Using FYI as an instrumented design probe, we conducted an exploratory study to examine how participants fact-check data claims when multiple AI and manual verification tools are simultaneously available. 
The complete study materials, including the article, dataset, and post-study interview questions, are provided in the supplementary materials.

\subsection{Participants}

We recruited 24 participants through university mailing lists and research group networks. Two were excluded due to data loss (one missing session logs, one missing session recording), yielding a final sample of 22 (14 male, 8 female; 12 aged 18--24, 10 aged 25--34). Participants held Bachelor's (10), Master's (5), or PhD (7) degrees from disciplinary backgrounds including computer science, HCI, data science, and engineering.
Self-reported prior experience with data visualization was high ($M = 5.68/7$, $SD = 1.09$), and daily use of generative AI tools was near the ceiling ($M = 6.18/7$, $SD = 1.14$). Participants reported moderate fact-checking experience ($M = 4.18/7$, $SD = 1.44$), while familiarity with the article's topic was mixed ($M = 3.68/7$, $SD = 1.94$).
The study was conducted entirely at Holistics Software, where the lead author was employed at the time and where research of this kind did not require IRB approval. Informed consent was obtained from each participant before the study.
Each participant was compensated USD~10 for their time.

\subsection{Materials}

We adopted a data-first approach, selecting the dataset before drafting the article to ensure experimental control.

\bpstart{Dataset.}
A movie industry dataset containing 1,724 films (1995--2018) with eight attributes (title, worldwide gross, production budget, release year, content rating, running time, genre, and IMDb rating) was constructed from publicly available sources~\cite{narechania2021nl4dv}. The movie domain was chosen for broad accessibility without requiring specialized knowledge, while the attributes support diverse verification operations (filtering, aggregation, cross-group comparison, and correlation).

\bpstart{Article.}
Grounded in the dataset, we authored a movie industry analysis article covering box-office performance, ratings, and genre patterns, designed to resemble typical data-driven journalism. We embedded six data claims that span the three verification outcomes (3 verified, 2 misleading, and 1 unverifiable). We further designed the claims so that no single tool would suffice to check them: some can be settled by a direct data operation such as filtering or charting a value, others require multi-step analysis such as comparing groups or computing a correlation, and others demand judgment beyond the data, such as recognizing a statistically true but rhetorically overreaching statement or a claim the dataset cannot settle. This variation was intended to require participants to combine multiple verification strategies rather than rely on any single tool or pathway.

\subsection{Procedure}

Each session lasted approximately 60 minutes and was conducted remotely via Zoom. The study followed three phases.

\bpstart{Phase~I: Onboarding and warm-up (20 min).}
Participants received a brief introduction to data claims and a guided walkthrough of \app's interface, followed by a warm-up task to reach a baseline level of proficiency with each tool before beginning the main task. They were also told that the study focused on understanding their verification process rather than evaluating their speed or correctness.

\bpstart{Phase~II: Main task (30 min).}
Participants were presented with the movie article and its grounding dataset. They were first instructed to read the article and identify data claims they considered worth checking, either by running AI detection, manually highlighting passages, or both. Then they freely investigated those claims using any combination of \app's four tools. We intentionally did not prescribe a fixed claim order or tool order, because the design-probe study aimed to observe how participants appropriated tools, formed workflows, and calibrated trust during verification.
Because interaction logs alone cannot capture the reasoning behind tool choices or trust judgments, participants followed a concurrent think-aloud protocol~\cite{ericsson1993protocol}, verbalizing their reasoning, hypotheses, and motivations for switching between tools throughout the task. 
For each investigated claim, they submitted a verdict (verified, misleading, or unverifiable) along with a confidence rating (1--7)  and a ranked list of which tools they found helpful. For misleading or unverifiable claims, they additionally provided a severity rating and a short explanation or correction.

\bpstart{Phase~III: Post-study questionnaire and interview (10 min).}
After the main task, participants completed a questionnaire covering demographics, prior experience, perceived helpfulness of each tool (7-point Likert), the NASA Task Load Index (NASA-TLX; 7-point scale, six dimensions), and the System Usability Scale (SUS; 10 items, 5-point scale). The session ended with a semi-structured interview probing four topics: claim detection strategy, verification strategy, tool switching triggers, and tool utility with AI trust preferences.

\subsection{Data Collection and Analysis}

\app logged all user interactions as timestamped events, capturing 2,250 events across 25 event types. From these logs, we derived a per-participant metrics file (22 rows, 64 columns) aggregating session-level measures with questionnaire responses, and a master event table (2,250 rows) preserving all raw interactions.

For qualitative analysis, think-aloud protocols and interview transcripts were analyzed using thematic analysis~\cite{braun2006using}. Three researchers independently read all transcripts, produced per-participant memos, and iteratively developed themes through constant comparison across three coding passes. Themes were then discussed and consolidated to reach consensus. The resulting themes are integrated with quantitative findings in \cref{sec:results}.

\section{Results}
\label{sec:results}

\begin{table}[t]
  \centering
  \caption{Per-participant summary (P03 and P10 excluded due to data loss; see \cref{sec:study}). Claims = highlighted claims; Verdicts = submitted verdicts; Confidence = mean self-reported confidence (1--7); Duration = session length in minutes.}
  \label{tab:participant-summary}
  \resizebox{\columnwidth}{!}{%
  \begin{tabular}{l r r r r r r}
    \toprule
    \textbf{ID} & \textbf{Claims} & \textbf{Verdicts} & \textbf{Verified} & \textbf{Misleading} & \textbf{Confidence} & \textbf{Duration} \\
    \midrule
    P01 &  5 &  5 & 3 & 0 & 5.20 & 21.7 \\
    P02 & 15 & 11 & 6 & 2 & 6.45 & 26.3 \\
    P04 & 12 &  6 & 4 & 0 & 6.17 & 28.4 \\
    P05 &  6 &  6 & 3 & 2 & 7.00 & 24.8 \\
    P06 &  8 &  5 & 4 & 1 & 6.40 & 25.8 \\
    P07 &  5 &  5 & 4 & 1 & 6.80 & 33.3 \\
    P08 &  4 &  4 & 4 & 0 & 7.00 & 27.1 \\
    P09 &  9 &  7 & 5 & 0 & 6.43 & 30.9 \\
    P11 &  6 &  6 & 6 & 0 & 6.67 & 22.0 \\
    P12 &  4 &  4 & 4 & 0 & 7.00 & 12.5 \\
    P13 &  5 &  5 & 3 & 2 & 6.80 & 23.8 \\
    P14 &  6 &  6 & 3 & 2 & 5.33 & 25.9 \\
    P15 & 14 &  9 & 7 & 1 & 6.33 & 22.5 \\
    P16 &  6 &  6 & 3 & 3 & 6.67 & 18.9 \\
    P17 &  7 &  6 & 4 & 1 & 4.67 & 24.9 \\
    P18 &  6 &  6 & 5 & 0 & 6.00 & 15.8 \\
    P19 &  6 &  6 & 5 & 1 & 5.83 & 27.7 \\
    P20 & 13 &  7 & 5 & 2 & 6.57 & 30.4 \\
    P21 &  7 &  7 & 5 & 1 & 6.00 & 22.0 \\
    P22 &  5 &  5 & 4 & 1 & 6.60 & 26.1 \\
    P23 &  7 &  6 & 4 & 2 & 6.83 & 17.6 \\
    P24 &  5 &  5 & 4 & 1 & 7.00 & 26.3 \\
    \midrule
    \textbf{Mean} & 7.3 & 6.0 & 4.3 & 1.0 & 6.35 & 24.3 \\
    \bottomrule
  \end{tabular}%
  }
\end{table}

The following analysis combines interaction logs, self-reported metrics, think-aloud protocols, and post-study interviews. We organize findings around the four research questions, preceded by a behavioral overview. \Cref{tab:themes} summarizes the qualitative themes identified through thematic analysis, which structure the narrative in each subsection.

\begin{table}[t]
  \centering
  \caption{Tool usage summary. Claims = claims where the tool was used at least once; Users = participants who used the tool; Most Helpful = percentage of verdicts selecting this tool as most helpful; Helpfulness = post-task rating (1--7 Likert).}
  \label{tab:tool-usage}
  \resizebox{\columnwidth}{!}{%
  \begin{tabular}{lrrrr}
    \toprule
    \textbf{Tool} & \textbf{Claims} & \textbf{Users} & \textbf{Most Helpful} & \textbf{Helpfulness} \\
    \midrule
    \toolAutoCheckT  &  90 (57.7\%) & 18 (82\%) & 22.6\% & 5.59 ($\pm$1.05) \\
    \toolAIChatT     &  91 (58.3\%) & 21 (95\%) & 35.3\% & 5.14 ($\pm$1.55) \\
    \toolChartT      & 105 (67.3\%) & 22 (100\%) & 30.1\% & 5.59 ($\pm$1.50) \\
    \toolTableT      &  45 (28.8\%) & 21 (95\%) & 12.0\% & 5.27 ($\pm$1.28) \\
    \bottomrule
  \end{tabular}%
  }
\end{table}

\subsection{Overview}

Participants highlighted 161 claims across all sessions (88 AI-detected, 73 manually added). Five were dismissed, leaving 156 actively investigated claims.
Across the 22 participants, 133 verdicts were submitted, 95 verified (71.4\%), 23 misleading (17.3\%), and 15 unverifiable (11.3\%). These counts describe how broadly participants chose to investigate the article. They should not be read as evidence that finding more claims is always better, since exhaustive verification may increase workload in real reading contexts.

Session durations ranged from 12.5 to 33.3 minutes ($M = 24.3$, $SD = 5.0$), with participants highlighting 7.3 claims on average ($SD = 3.2$) and submitting 6.0 verdicts ($SD = 1.6$).
Individual differences were pronounced. The number of verdicts ranged from 4 (P08, P12) to 11 (P02), and P08 and P12 investigated only AI-detected claims, while P02 and P15 manually added numerous claims (\cref{tab:participant-summary}).

System usability was rated ``good'' (SUS $M = 74.1$, $SD = 13.4$) with moderate cognitive load (NASA-TLX $M = 3.1/7$, $SD = 0.7$), suggesting that the multi-tool design was feasible without overwhelming participants.

\subsection{RQ1: How Do Users Detect Data Claims?}
\label{sec:results-detection}

FYI offered two ways to surface claims: AI auto-detection and manual highlighting. Participants differed mainly in which pathway they began and why.

\bpstart{AI-first detection} (15/22; \cref{tab:themes}).
The majority of participants used AI detection as an efficient starting point, then supplemented it with manual scanning.
As P01 noted, ``\textit{I was lazy, I was reluctant to pick the claims one by one.}''
However, few treated AI detection as sufficient. For instance, P11 let AI find four claims, then manually added two more.
Trust in AI detection was moderate even within this group. P09 observed that AI ``\textit{always detects a very, very long sentence}'' rather than an accurate claim boundary, requiring manual adjustment.

\bpstart{Manual-first detection} (7/22; \cref{tab:themes}).
A smaller group systematically scanned for quantitative cues such as numbers, superlatives, and comparators before running AI as a check.
P02 explained, ``\textit{I don't want to be too dependent on using the AI, so first I just do it manually.}''
Data-savvy participants (P02, P04, P15) consistently preferred this approach, with P15 finding manual highlighting faster than waiting for AI to process the article.

Overall, AI-first detection was more common, but participants rarely treated AI detection as complete. Manual-first participants scanned by themselves to maintain control before optionally using AI as a check.
\subsection{RQ2: How Do Users Verify Data Claims?}
\label{sec:results-verification}

All four verification tools achieved broad adoption (\cref{tab:tool-usage}).
On average, participants used 3.7 out of 4 tools ($SD = 0.48$, range: 3--4).
When measuring per-claim adoption (a tool being used at least once per investigated claim), \toolChart was used most frequently (105 claims, 67.3\%), followed by \toolAIChat (91 claims, 58.3\%) and \toolAutoCheck (90 claims, 57.7\%), while \toolTable was used less frequently (45 claims, $28.8\%$).
Despite the frequency difference, post-task helpfulness ratings were uniformly positive ($M > 5.0/7$ for all tools).

\begin{figure}[t]
  \centering
  \includegraphics[width=\columnwidth, alt={One row per participant, 22 rows, ordered from participants who most often opened a claim with an AI tool at the top (P09, P06, P11, P16, P19) to those who most often opened manually at the bottom (P15, P18, P07, P04). Each row holds four to eight pills, and each pill holds two to nine markers. AI-first pills fill the upper two-thirds and manual-first pills the lower third, but neither is exclusive: P05, P02, P24, P14, and P20 alternate between the two within a single session, and P23 and P17 sit at the boundary with short pills of two or three markers throughout.}]{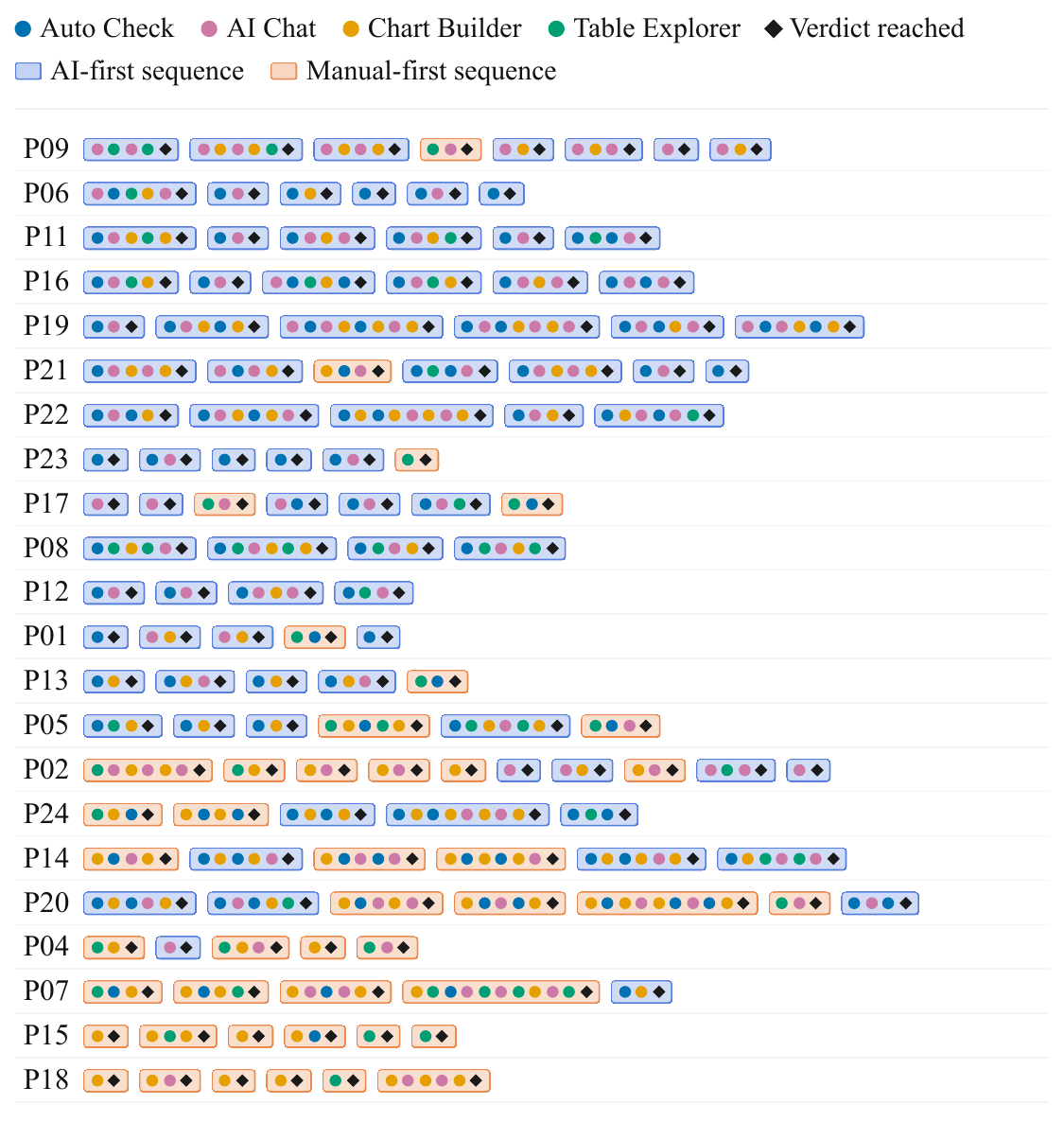}
  \caption{Tool-use sequences for every participant (across 22 participants, 131 claims). Each pill is one claim, tracing the ordered tools used to evaluate it from left to right and ending at the verdict ($\blacklozenge$); claims appear in the order they were addressed. Pill shading marks whether the claim opened with an AI tool (\toolAutoCheck/\toolAIChat) or a manual tool (\toolChart/\toolTable).}
  \label{fig:sequences}
\end{figure}

Think-aloud protocols and interaction logs suggested three recurring workflow orientations, though participants often shifted between them across claims instead of following a fixed strategy throughout the session (\cref{fig:sequences}).
The most common strategy was \emph{AI-first, manual confirmation} (9/22; \cref{tab:themes}).
These participants initiated their process with  \toolAutoCheck or \toolAIChat to establish a baseline verdict, subsequently verifying the findings through \toolChart or \toolTable.
As P12 explained, ``\textit{Auto Check and AI Chat have the same result\ldots~then I double-check with Chart Builder to make sure}.''
Conversely, a second group favored \emph{manual-first, AI as supplement} (6/22; \cref{tab:themes}).
P24 described this sequence: ``\textit{after I see the chart, I can have 90\% confidence that I have the correct answer. Then I will use Auto Check to finish the rest 10\%.}''
Notably, P15 never used AI for verification, entirely relying on a manual approach.
The third group engaged in \emph{parallel co-review} (4/22; \cref{tab:themes}), running AI and manual tools simultaneously.
P14 clarified this strategy as treating AI as a collaborator, ``\textit{set it on, like, a sub-agent kind of mode, and then do my own thing, and then converge to see if we come up with the same verdict.}''

\begin{figure}[t]
  \centering
  \includegraphics[width=\columnwidth, alt={Four labeled nodes in a rectangle, Auto Check upper left, AI Chat upper right, Table Explorer lower left, Chart Builder lower right, joined by twelve curved arrows carrying counts. The heaviest arrows are Auto Check to AI Chat with 49, AI Chat to Chart Builder with 44, Chart Builder to AI Chat with 38, and Auto Check to Chart Builder with 37, so the densest traffic runs along the top edge and down the right side. Every arrow touching Table Explorer is thin, ranging from 9 to 18, leaving the lower left corner sparse.}]{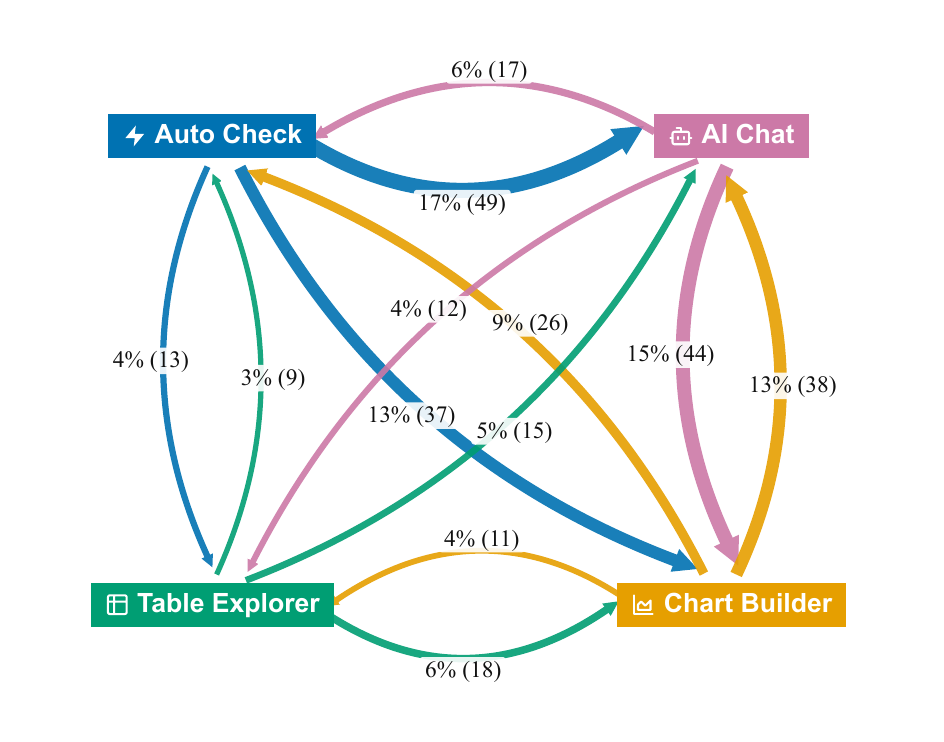}
  \caption{Transitions between tools (289 transitions across 131 claims). Arrows run from the previously used tool to the next; line thickness encodes each transition's share of all transitions.}
  \label{fig:tool-transitions}
\end{figure}

More notably, these archetypes were not fixed types but tendencies. Participants adapted their approach across claims as system reliability and familiarity shifted, often moving between archetypes within a single session.
For instance, P13 initially relied on an AI-first approach but pivoted to a \toolChart-dominant approach after \toolAutoCheck gave an incorrect verdict.
By contrast, P24 began with manual exploration but gradually incorporated \toolAutoCheck as a background process as their comfort level increased.
The tool-transition graph (\cref{fig:tool-transitions}) reflects this fluidity. Rather than following a fixed order, participants moved frequently among tools. The most common single transition was \toolAutoCheck$\rightarrow$\toolAIChat, while \toolAIChat and \toolChart formed the busiest two-way exchange. \toolTable was used least overall, as it provided limited support for aggregation-focused claims.

Overall, tool transitions were primarily driven by a desire for triangulation and confidence-building rather than dissatisfaction with the initial tool.
As P21 noted, ``\textit{It's not abandoning, it's more about reinforcing the claim.}''

\begin{table*}[t]
  \centering
  \caption{Qualitative themes from thematic analysis of think-aloud protocols and post-study interviews ($N = 22$). Themes are not mutually exclusive; participants may appear in multiple themes. PIDs enable cross-referencing with \cref{tab:participant-summary}.}
  \label{tab:themes}
  \resizebox{\textwidth}{!}{%
  \begin{tabular}{l l l r l}
    \toprule
    \textbf{RQ} & \textbf{ID} & \textbf{Theme} & \textbf{n} & \textbf{Participants} \\
    \midrule
    \multirow{2}{*}{RQ1: Detection}
      & T1 & AI-first detection, manual supplement & 15 & P01, P05, P06, P07, P08, P09, P11, P12, P16, P17, P19, P21, P22, P23, P24 \\
      & T2 & Manual-first detection, AI as recheck & 7 & P02, P04, P13, P14, P15, P18, P20 \\
    \midrule
    \multirow{4}{*}{RQ2: Verification}
      & T3 & AI-first workflow, manual confirmation & 9 & P01, P08, P11, P12, P16, P17, P19, P21, P23 \\
      & T4 & Manual-first workflow, AI as supplement & 6 & P02, P04, P15, P18, P20, P24 \\
      & T5 & Parallel co-review & 4 & P07, P14, P19, P22 \\
    \midrule
    \multirow{3}{*}{RQ3: Verdicts}
      & T7 & Visual evidence as stopping criterion & 10 & P01, P02, P05, P07, P13, P14, P19, P20, P22, P24 \\
      & T8 & Multi-tool convergence as confidence signal & 10 & P05, P08, P12, P14, P17, P19, P20, P21, P22, P24 \\
      & T9 & Claim decomposition and statistical reasoning & 15 & P02, P04, P05, P07, P09, P12, P13, P14, P17, P18, P19, P20, P22, P23, P24 \\
    \midrule
    \multirow{3}{*}{RQ4: Trust}
      & T10 & Trust builds through convergence, erodes through inconsistency & 12 & P05, P07, P08, P12, P13, P14, P17, P19, P20, P21, P22, P24 \\
      & T11 & Transparency modulates trust & 6 & P06, P17, P20, P21, P22, P24 \\
      & T12 & AI initiates, human decides (dominant reliance model) & 13 & P01, P05, P07, P08, P09, P11, P12, P14, P17, P18, P20, P21, P22 \\
    \bottomrule
  \end{tabular}%
  }
\end{table*}

\subsection{RQ3: How Do Users Reach Verdicts?}
\label{sec:results-verdicts}

Participants reported consistently high confidence across all verdict types ($M = 6.35/7$, $SD = 1.20$), spending an average of 33.6 seconds ($SD = 15.2$) on each verdict.
Confidence was comparably high for verified ($M = 6.6$) and misleading ($M = 6.3$) claims, but notably lower for unverifiable ones ($M = 4.8$, $SD = 1.78$), reflecting the inherent uncertainty of that judgment.
Thematic analysis identified three mechanisms driving participants' final determinations. 

\bpstart{Visual evidence as stopping criterion} (10/22; \cref{tab:themes}).
For these participants, a self-constructed chart in \toolChart that clearly aligned with or debunked a claim was often decisive, at times overriding the other tools' conclusions.
In 55 instances, users deliberately launched \toolChart \emph{after} \toolAutoCheck completed, using charts to audit AI conclusions.
P02 noted, ``\textit{If the chart already supports or dismisses the claim, then I can confidently say that I've already verified it.}''
However, this created a usability--confidence paradox. Participants with lower visualization literacy (e.g., P17, who found charting "\textit{too much of a mental task,}" avoided \toolChart entirely) were forced to rely on AI tools they trusted less.

\bpstart{Multi-tool convergence} (10/22; \cref{tab:themes}).
In contrast, these participants treated no single tool as sufficient. They stopped only once multiple independent tools \emph{agreed}, so a divergence among tools meant continuing to investigate rather than concluding.
P19 illustrated this layered confidence as ``\textit{If both these agents give me the correct answer, I'm mostly sure it is correct. If additionally I'm able to plot it on the chart by myself, then it is 100\% correct}.''

\bpstart{Claim decomposition and statistical reasoning} (15/22; \cref{tab:themes}).
Reaching a verdict required substantive analytical reasoning beyond tool operation. 
Participants actively decomposed complex sentences into testable sub-claims (P17: "\textit{there are two claims within the claim}") and debated statistical nuances like correlation versus causation or mean versus median.
Furthermore, they successfully recognized when claims required external context beyond the provided dataset, correctly labeling them as "unverifiable."

\subsection{RQ4: How Do Users Calibrate Trust between AI and Manual Tools?}
\label{sec:results-trust}

\begin{figure}[b]
  \centering
  \includegraphics[width=\columnwidth, alt={Two three-by-three matrices side by side. Auto Check, left, reading rows Verified, Misleading, Unverifiable against columns V, M, U: 44, 0, 1; then 17, 12, 0; then 3, 2, 0. Participant, right, in the same order: 42, 1, 2; then 14, 15, 0; then 1, 1, 3. Both agree on almost every Verified claim. They differ on the other two rows: Auto Check calls 17 of the 29 Misleading claims Verified against the participants' 14, and Auto Check never returns Unverifiable at all, while participants use it 3 times and match the researcher on all 3.}]{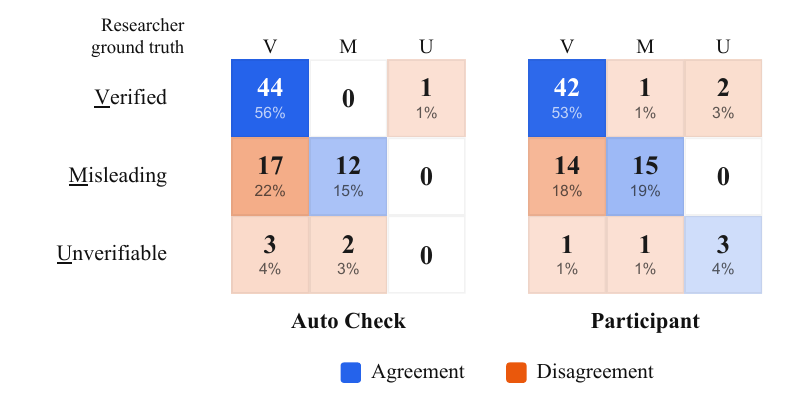}
  \caption{Verdict agreement with researcher ground truth ($n=79$). Rows give the researcher's ground-truth verdict; columns give the verdict from \toolAutoCheck (left) and the participant (right), where V~=~Verified, M~=~Misleading, U~=~Unverifiable. Diagonal cells (blue) are agreements and off-diagonal cells (orange) are disagreements; each cell reports the count and its share of all 79 verdict instances, and shade encodes magnitude.}
  \label{fig:verdict-accuracy}
\end{figure}

To study how participants calibrated their trust, we first examine how accurate \toolAutoCheck's verdicts were. We compared the researchers' verdicts on the six embedded claims against outputs from \toolAutoCheck and the participants, and this analysis covers only the 79 verdict instances for which all three labels were available (\cref{fig:verdict-accuracy}).
Overall, \toolAutoCheck and participants matched the researchers' labels at similar rates of 71\% and 76\%, respectively, yet performance varies by claim type. 
For \emph{verified} claims, they both aligned with the researchers' verdicts nearly universally. \toolAutoCheck correctly assessed 44 of 45 verdicts and participants 42 of 45. 
While on the \emph{misleading} claims, alignment was weaker for both: \toolAutoCheck matched the \emph{misleading} label on 41\% of these verdicts (12 of 29) and participants on 52\% (15 of 29), with the rest mostly misclassified as \emph{verified}. Most misjudgments came from one specific claim that takes a factual statistic that dramas hold the highest average IMDb rating and twists it into a misleading causal inference that audiences \emph{universally prefer serious storytelling}.

The clearest disagreement appeared on the single \emph{unverifiable} claim, which cannot be proven or disproven using the dataset alone. Across 79 claims, \toolAutoCheck returned an \emph{unverifiable} verdict only once, but never on the unverifiable claim defined by researchers. Instead, it consistently categorized it as \emph{verified} or \emph{misleading}.
By contrast, participants labeled this claim as \emph{unverifiable} 3 times out of 5 total verdicts they gave. 
On the single unverifiable claim, \toolAutoCheck never matched the researchers' label, whereas participants did so in 3 of 5 cases.
This pattern points to a possible blind spot: \toolAutoCheck was accurate on the straightforward, data-answerable claims but did not flag the unverifiable claim that required contextual judgment.
Participants mitigated this limitation by using AI for initial analysis while retaining final judgments for themselves. We further analyze their trust-calibration pattern as follows.

\bpstart{Convergence builds trust, inconsistency erodes it} (12/22; \cref{tab:themes}).
Trust calibrated dynamically with experience.
Alignment between AI and manual tools boosted confidence, while numerical inconsistencies, even with directional agreement, quickly eroded it. 
For example, P17 found different Pearson correlations from \toolAutoCheck and \toolAIChat: ``\textit{why are the numbers not consistent?}''
Moreover, first impressions proved critical. P13 abandoned \toolAutoCheck for the rest of the session after an initial incorrect verdict.
Interestingly, participants evaluated different output modalities independently, often trusting \toolAutoCheck's textual conclusions while disregarding the broken or incorrectly scaled charts it frequently generated.

\bpstart{Transparency modulates trust} (6/22; \cref{tab:themes}).
Participants trusted AI more when its reasoning process was visible.
P20 preferred \toolAutoCheck because ``\textit{it shows the steps\ldots~so it feels more controllable},'' while P22 deeply inspected \toolAIChat's Python code to verify its underlying logic.
By exposing intermediate operations, the system could empower users to calibrate their trust based on comprehensible evidence rather than blind faith.

\bpstart{AI initiates, human decides} (13/22; \cref{tab:themes}).
The dominant reliance model positioned AI as an initial assessor, with humans retaining final decision-making authority. 
For instance, P18 said ``\textit{I prefer AI doing the initial. But not the end result.}''
However, the strictness of this human-in-the-loop behavior appeared to vary with participants' data literacy and professional background. Individuals with high data literacy, such as AI/data professionals (P04, P12, P15), actively minimized their use of AI to maintain strict analytical control. 
In contrast, participants with lower data literacy (e.g., P16, P23) were far more willing to surrender this control, leaning toward full AI delegation rather than independent verification.

\section{Discussion}
\label{sec:discussion}

Our findings suggest that navigating the detection--verification--determination pipeline is highly dynamic when multiple modalities are available simultaneously. 
We discuss what these behaviors reveal about multi-tool orchestration, human-AI complementarity, and trust dynamics, and distill these observations into design implications for future mixed-initiative fact-checking systems.

\subsection{The Pipeline in Practice}

Within our open-ended design-probe setting, participants rarely followed a strict detection--verification--determination sequence, and the boundaries between detection and verification largely dissolved.
Participants frequently discovered additional check-worthy claims \emph{during} the verification phase, prompting them to return to the text and manually highlight new statements. 
Verification itself proved deeply iterative. Participants actively decomposed complex claims, switched tools to test alternative interpretations, and occasionally revised earlier verdicts upon encountering contradictory evidence. 
This cyclical behavior aligns with sensemaking theory~\cite{pirolli2011introduction}, suggesting that in-situ fact-checking involves fluid movement between information foraging and synthesis rather than a rigid forward progression.

FYI's role as an instrumented design probe helped capture this process. 
By embedding multiple tools within the reading context and logging tool transitions, FYI made it possible to observe how participants constructed workflows across claims and modalities.
 Instead of following a fixed tool order, participants adapted their workflows based on emerging evidence, the analytic operation required by the claim, and their evolving trust in the system.

\subsection{Human-AI Complementarity and Visual Auditing}

A central insight from our study is that participants did not treat AI and manual tools as substitutes, but as complements serving distinct epistemic roles. 
The emergence of diverse workflow archetypes---\emph{AI-first with manual confirmation}, \emph{manual-first with AI supplement}, and \emph{parallel co-review}---reflects different strategies for distributing cognitive labor across the human-AI spectrum.

This multi-strategy behavior extends findings from interactive fact-checking systems such as WebSeek~\cite{huang2026facilitating} and DataDive~\cite{kim2024datadive}, where users combined extraction, tabular inspection, and chart construction. 
By introducing generative AI into the toolkit, \app created a richer space for strategy composition. 
Notably, no single workflow dominated, and strategies fluidly shifted within individual sessions. 
This heterogeneity suggests that verification systems should support flexible composition rather than prescribing a fixed optimal process.

Crucially, visualization emerged as a distinct \emph{verification modality}, not merely an auxiliary feature. 
Participants frequently launched \toolChart \emph{after} \toolAutoCheck completed, using self-built charts to audit AI conclusions. 
This pattern extends the emphasis-checking paradigm~\cite{choi2024emphasischecker} by revealing a deeper cognitive preference. 
Users actively used visualization tools as instruments of oversight. They explicitly valued the agency of constructing their own evidence over passively consuming AI-produced summaries.

\subsection{Trust Dynamics and the Usability-Confidence Paradox}

Our results reveal trust as a dynamic process that shifts with accumulated experience rather than a stable individual trait~\cite{yin2019understanding}. Trust was built through cross-tool convergence, but was easily eroded when outputs were numerically inconsistent even if directionally correct, and in extreme cases collapsed permanently after a single AI error. These dynamics extend prior work on trust calibration by showing that in multi-tool environments, users evaluate output modalities independently, often accepting AI’s textual conclusions while dismissing its flawed visual charts. A few participants chose to cross-validate one AI tool against another instead of referencing original data. They either used \toolAIChat to re-examine \toolAutoCheck (P11) or compared figures across the two AI tools (P17). However, such AI-on-AI checks could produce conflicting values that undermine user confidence in automated outputs, underscoring the value of auditing with self-built charts.

While high overall confidence and the similar ground-truth agreement patterns of \toolAutoCheck and participants (\cref{fig:verdict-accuracy}) raise valid concerns regarding automation bias and anchoring effects, the \emph{AI initiates, human decides} reliance model shows users actively attempting to maintain final authority. However, this resistance is gated by data literacy, resulting in a critical usability-confidence paradox. Participants who found chart authoring too cognitively demanding were forced to rely on automated AI tools that they otherwise trusted less. This suggests a vulnerability that limited visualization literacy may constrain a user's ability to audit algorithmic outputs. Furthermore, we found that process visibility, such as inspecting generated code, acted as a modulator of trust and was associated with more selective override behaviors.

\subsection{Design Implications}

Based on these empirical insights, we propose four implications for future mixed-initiative fact-checking systems.

\bpstart{DI\textsubscript{1}: Design AI as an initial guide, not a definitive authority.}
The dominant \emph{AI initiates, human decides} pattern suggests that AI is most useful when providing an investigative starting point.
Systems should present AI outputs as provisional hypotheses that actively invite human verification, instead of definitive verdicts that require significant cognitive effort to override. 

\bpstart{DI\textsubscript{2}: Elevate visualization as a core auditing modality.}
Given its critical role in post-AI verification, user-constructed visualization confers a sense of agency and interpretive confidence that AI-generated evidence may not readily substitute.
Fact-checking systems should integrate visualization authoring as a primary capability, and consider AI-assisted chart suggestion to lower the entry barrier for users with limited visualization literacy.

\bpstart{DI\textsubscript{3}: Support flexible, cross-modal workflow composition.}
The fluidity of the three observed workflow archetypes argues against rigid, step-by-step verification wizards.
Systems should provide independent, composable tools that users can freely arrange and interleave according to the type of evidence required, the amount of analysis involved, and their own expertise.

\bpstart{DI\textsubscript{4}: Make AI reasoning and uncertainty visible.}
Because process transparency appears to modulate trust calibration, future systems must move beyond opaque final outputs. 
Systems should expose intermediate reasoning, data queries, and underlying computation steps, alongside confidence scores, to support informed calibration over blind acceptance or algorithmic aversion.

\section{Limitations and Future Work}
\label{sec:limitations}

While our design probe provides rich behavioral insights, several factors constrain the scope of our findings.

\bpstart{Dataset availability and ecological scope.} 
FYI assumes access to a relevant structured dataset, which allowed us to focus on fact-checking against inspectable data. This assumption fits data-driven articles that disclose their underlying data~\cite{zamith2019transparency}. Yet it excludes articles that make quantitative claims without releasing the underlying data, or cases where the available data are incomplete, messy, or require substantial cleaning before verification. Future systems should support upstream data discovery, provenance assessment, and data preparation in addition to the dataset-grounded fact-checking workflow studied here.

\bpstart{Domain and sample.}
The study used a single article in the movie domain with one accompanying dataset. Verification strategies may differ substantially in domains such as finance or public health, where claims involve multi-table joins, real-time data, or specialized domain knowledge. Our university-affiliated participants reported high visualization literacy and AI familiarity, which may limit generalizability to readers with lower data or visualization literacy. The observed workflow diversity may therefore underestimate the challenges faced by less experienced users.

\bpstart{Accuracy and ground truth.}
Ground truth in our study was limited to the six embedded data claims, against which we report \toolAutoCheck's and participants' verdict accuracy (\cref{sec:results-trust}). The remaining 75 claims were surfaced by participants and lay outside this set; many are not data claims, and none carry researcher veracity labels, as they were never intended as study ground truth. Future studies could establish ground truth for every claim in an article and examine textual and data claims in conjunction, enabling a fuller assessment of verification accuracy across claim types.

\bpstart{System and model dependence.}
The current \toolTable supports direct lookup and filtering, and future versions could add lightweight in-table analytical features, such as conditional formatting or simple grouped summaries, to reduce the need to switch tools for basic comparisons. In addition, all AI-powered components rely on GPT-4.1, whose outputs vary with model version, prompt design, and stochastic sampling. Known limitations in LLM numerical reasoning~\cite{yuan2023large} and hallucination tendencies~\cite{pesaranghader2026hallucination} affected participants' trust dynamics (e.g., \toolAutoCheck producing broken charts). The multi-tool design partially mitigates this by providing non-AI verification paths through \toolTable and \toolChart, but findings may not generalize to other model families or future model versions.

\section{Conclusion}
\label{sec:conclusion}

\app is an in-situ browser extension which spans the human-AI spectrum from fully automated verification (\toolAutoCheck) through user-directed AI assistance (\toolAIChat) to unmediated data exploration (\toolTable, \toolChart). We used it as a design probe for an exploratory study ($N = 22$). Our findings revealed that participants did not follow a fixed linear pipeline. Instead, they constructed heterogeneous workflows, including AI-first with manual confirmation, manual-first with AI supplement, and parallel co-review.
Visualization served as a primary auditing mechanism, as participants used self-built charts to inspect and challenge AI conclusions.
Trust was calibrated dynamically based on cross-tool convergence, output inconsistency, and process transparency.
These findings suggest that future fact-checking environments could use AI as an initial guide rather than a definitive authority, since participants tended to reach for it first and found it broadly accurate, while treating manual tools such as visualization as indispensable rather than secondary. This points toward keeping both humans and AI in the loop, prioritizing visualization as a core verification capability, and supporting flexible cross-modal workflows.

\section*{Supplemental Materials}
\label{sec:supplemental_materials}

To facilitate subsequent studies on fact-checking tools, we open-source the \app software prototype at \url{https://github.com/DataVisards/FYI}. Supplemental material is available in the IEEE Xplore digital repository and includes (1) a video demonstration of \app outlining its claim detection, multi-tool verification, and verdict determination workflows; (2) the study article and movie dataset used in the evaluation, along with the researcher ground-truth labels for the six embedded claims, the study protocol, the interview guide, and the questionnaire instrument; (3) all LLM prompts for claim detection, \toolAutoCheck, and \toolAIChat; (4) the raw per-participant interaction logs with a data dictionary; (5) the per-participant think-aloud transcripts; and (6) the post-task questionnaire responses.

\acknowledgments{%
The authors used Claude Code (Anthropic) to help write and debug portions of the \app prototype's source code (\cref{sec:system-design}); all AI-generated code was reviewed and tested by the authors. Generative AI was not used for the study design, data, or analysis. We thank the members of the DataVisards Lab at HKUST, user study participants, and anonymous reviewers for their feedback during various stages of this work.
}

\bibliographystyle{abbrv-doi-hyperref}

\bibliography{main}

\begin{thebibliography}{10}

\bibitem{buholayka2024reference}
F.~Aljamaan, M.-H. Temsah, I.~Altamimi, A.~Al-Eyadhy, A.~Jamal, K.~Alhasan et al.
\newblock Reference hallucination score for medical artificial intelligence chatbots: Development and usability study.
\newblock {\em JMIR Med. Inform.}, 12,  art. no. e54345, Jul. 2024. \href{https://doi.org/10.2196/54345}
{doi: {{%
10\hspace{.1pt}\discretionary{.}{%
}{.}\hspace{.4pt}2196\discretionary{/}{%
}{/}54345}}}


\bibitem{lane2014effect}
P.~M. Allen, J.~A. Edwards, F.~J. Snyder, K.~A. Makinson, and D.~M. Hamby.
\newblock The effect of cognitive load on decision making with graphically displayed uncertainty information.
\newblock {\em Risk Anal.}, 34(8):1495--1505, Aug. 2014. \href{https://doi.org/10.1111/risa.12161}
{doi: {{%
10\hspace{.1pt}\discretionary{.}{%
}{.}\hspace{.4pt}1111\discretionary{/}{%
}{/}risa\hspace{.1pt}\discretionary{.}{%
}{.}\hspace{.4pt}12161}}}


\bibitem{aslett2024online}
K.~Aslett, Z.~Sanderson, W.~Godel, N.~Persily, J.~Nagler, and J.~A. Tucker.
\newblock Online searches to evaluate misinformation can increase its perceived veracity.
\newblock {\em Nature}, 625:548--556, Jan. 2024. \href{https://doi.org/10.1038/s41586-023-06883-y}
{doi: {{%
10\hspace{.1pt}\discretionary{.}{%
}{.}\hspace{.4pt}1038\discretionary{/}{%
}{/}s41586\discretionary{%
}{-}{-}023\discretionary{%
}{-}{-}06883\discretionary{%
}{-}{-}y}}}


\bibitem{balalau2022statcheck}
O.~Balalau, S.~Ebel, T.~Galizzi, I.~Manolescu, Q.~Massonnat, A.~Deiana et al.
\newblock Statistical claim checking: {StatCheck} in action.
\newblock In {\em Proc.\ {ACM} Int. Conf. Inf. \& Knowl. Manage. ({CIKM})}, pp. 4798--4802. {ACM}, New York, Oct. 2022. \href{https://doi.org/10.1145/3511808.3557198}
{doi: {{%
10\hspace{.1pt}\discretionary{.}{%
}{.}\hspace{.4pt}1145\discretionary{/}{%
}{/}3511808\hspace{.1pt}\discretionary{.}{%
}{.}\hspace{.4pt}3557198}}}


\bibitem{block2022influence}
J.~E. Block, S.~Esmaeili, E.~D. Ragan, J.~R. Goodall, and G.~D. Richardson.
\newblock The influence of visual provenance representations on strategies in a collaborative hand-off data analysis scenario.
\newblock {\em IEEE Trans. Visual. Comput. Graphics}, 29(1):1113--1123, Jan. 2023. \href{https://doi.org/10.1109/TVCG.2022.3209495}
{doi: {{%
10\hspace{.1pt}\discretionary{.}{%
}{.}\hspace{.4pt}1109\discretionary{/}{%
}{/}TVCG\hspace{.1pt}\discretionary{.}{%
}{.}\hspace{.4pt}2022\hspace{.1pt}\discretionary{.}{%
}{.}\hspace{.4pt}3209495}}}


\bibitem{botnevik2020brenda}
B.~Botnevik, E.~Sakariassen, and V.~Setty.
\newblock {BRENDA}: Browser extension for fake news detection.
\newblock In {\em Proc.\ {ACM} Int. Conf. Res. Develop. Inf. Retrieval ({SIGIR})}, SIGIR '20, pp. 2117--2120. {ACM}, New York, Jul. 2020. \href{https://doi.org/10.1145/3397271.3401396}
{doi: {{%
10\hspace{.1pt}\discretionary{.}{%
}{.}\hspace{.4pt}1145\discretionary{/}{%
}{/}3397271\hspace{.1pt}\discretionary{.}{%
}{.}\hspace{.4pt}3401396}}}


\bibitem{braun2006using}
V.~Braun and V.~Clarke.
\newblock Using thematic analysis in psychology.
\newblock {\em Qual. Res. Psychol.}, 3(2):77--101, Jan. 2006. \href{https://doi.org/10.1191/1478088706qp063oa}
{doi: {{%
10\hspace{.1pt}\discretionary{.}{%
}{.}\hspace{.4pt}1191\discretionary{/}{%
}{/}1478088706qp063oa}}}


\bibitem{yin2019understanding}
S.~Cao and C.-M. Huang.
\newblock Understanding user reliance on {AI} in assisted decision-making.
\newblock {\em Proc. ACM Hum.-Comput. Interact.}, 6(CSCW2),  art. no. 471,  23 pp., Nov. 2022. \href{https://doi.org/10.1145/3555572}
{doi: {{%
10\hspace{.1pt}\discretionary{.}{%
}{.}\hspace{.4pt}1145\discretionary{/}{%
}{/}3555572}}}


\bibitem{boldova2024cognitive}
B.~G. d.~S. Cezar and A.~C.~G. Ma{\c{c}}ada.
\newblock Cognitive overload, anxiety, cognitive fatigue, avoidance behavior and data literacy in big data environments.
\newblock {\em Inf. Process. \& Manage.}, 60(6),  art. no. 103482, Nov. 2023. \href{https://doi.org/10.1016/j.ipm.2023.103482}
{doi: {{%
10\hspace{.1pt}\discretionary{.}{%
}{.}\hspace{.4pt}1016\discretionary{/}{%
}{/}j\hspace{.1pt}\discretionary{.}{%
}{.}\hspace{.4pt}ipm\hspace{.1pt}\discretionary{.}{%
}{.}\hspace{.4pt}2023\hspace{.1pt}\discretionary{.}{%
}{.}\hspace{.4pt}103482}}}


\bibitem{chae2024perceiving}
J.~H. Chae and D.~Tewksbury.
\newblock Perceiving {AI} intervention does not compromise the persuasive effect of fact-checking.
\newblock {\em New Media Soc.}, 28(1):191--211, Jan. 2026. \href{https://doi.org/10.1177/14614448241286881}
{doi: {{%
10\hspace{.1pt}\discretionary{.}{%
}{.}\hspace{.4pt}1177\discretionary{/}{%
}{/}14614448241286881}}}


\bibitem{chegini2025repanda}
A.~Chegini, K.~Rezaei, H.~Eghbalzadeh, and S.~Feizi.
\newblock {R}e{P}anda: Pandas-powered tabular verification and reasoning.
\newblock In {\em Proc.\ 63rd Annu. Meeting Assoc. Comput. Linguistics ({ACL})}, pp. 32200--32212. Association for Computational Linguistics, Vienna, Austria, Jul. 2025. \href{https://doi.org/10.18653/v1/2025.acl-long.1549}
{doi: {{%
10\hspace{.1pt}\discretionary{.}{%
}{.}\hspace{.4pt}18653\discretionary{/}{%
}{/}v1\discretionary{/}{%
}{/}2025\hspace{.1pt}\discretionary{.}{%
}{.}\hspace{.4pt}acl\discretionary{%
}{-}{-}long\hspace{.1pt}\discretionary{.}{%
}{.}\hspace{.4pt}1549}}}


\bibitem{chen2020tabfact}
W.~Chen, H.~Wang, J.~Chen, Y.~Zhang, H.~Wang, S.~Li et al.
\newblock {TabFact}: A large-scale dataset for table-based fact verification.
\newblock In {\em Proc.\ Int. Conf. Learn. Representations ({ICLR})}. Virtual Conference, Jun. 2020. \href{https://doi.org/10.48550/arXiv.1909.02164}
{doi: {{%
10\hspace{.1pt}\discretionary{.}{%
}{.}\hspace{.4pt}48550\discretionary{/}{%
}{/}arXiv\hspace{.1pt}\discretionary{.}{%
}{.}\hspace{.4pt}1909\hspace{.1pt}\discretionary{.}{%
}{.}\hspace{.4pt}02164}}}


\bibitem{chen2022crossdata}
Z.-T. Chen and H.~Xia.
\newblock {CrossData}: Leveraging text-data connections for authoring data documents.
\newblock In {\em Proc.\ {ACM} {CHI} Conf. Human Factors Comput. Syst. ({CHI})}, CHI '22,  art. no. 95. {ACM}, New York, Apr. 2022. \href{https://doi.org/10.1145/3491102.3517485}
{doi: {{%
10\hspace{.1pt}\discretionary{.}{%
}{.}\hspace{.4pt}1145\discretionary{/}{%
}{/}3491102\hspace{.1pt}\discretionary{.}{%
}{.}\hspace{.4pt}3517485}}}


\bibitem{cheng2023binder}
Z.~Cheng, T.~Xie, P.~Shi, C.~Li, R.~Nadkarni, Y.~Hu et al.
\newblock Binding language models in symbolic languages.
\newblock In {\em Proc.\ Int. Conf. Learn. Representations ({ICLR})}. Kigali, Rwanda, May 2023. \href{https://doi.org/10.48550/arXiv.2210.02875}
{doi: {{%
10\hspace{.1pt}\discretionary{.}{%
}{.}\hspace{.4pt}48550\discretionary{/}{%
}{/}arXiv\hspace{.1pt}\discretionary{.}{%
}{.}\hspace{.4pt}2210\hspace{.1pt}\discretionary{.}{%
}{.}\hspace{.4pt}02875}}}


\bibitem{das2025misvisfix}
A.~K. Das and K.~Mueller.
\newblock {MisVisFix}: An interactive dashboard for detecting, explaining, and correcting misleading visualizations using large language models.
\newblock {\em IEEE Trans. Visual. Comput. Graphics}, 32(1):134--144, Jan. 2026. \href{https://doi.org/10.1109/TVCG.2025.3633884}
{doi: {{%
10\hspace{.1pt}\discretionary{.}{%
}{.}\hspace{.4pt}1109\discretionary{/}{%
}{/}TVCG\hspace{.1pt}\discretionary{.}{%
}{.}\hspace{.4pt}2025\hspace{.1pt}\discretionary{.}{%
}{.}\hspace{.4pt}3633884}}}


\bibitem{deverna2024llmfactcheckbehavior}
M.~R. DeVerna, H.~Y. Yan, K.-C. Yang, and F.~Menczer.
\newblock Fact-checking information from large language models can decrease headline discernment.
\newblock {\em Proc. Natl. Acad. Sci. USA}, 121(50),  art. no. e2322823121, Dec. 2024. \href{https://doi.org/10.1073/pnas.2322823121}
{doi: {{%
10\hspace{.1pt}\discretionary{.}{%
}{.}\hspace{.4pt}1073\discretionary{/}{%
}{/}pnas\hspace{.1pt}\discretionary{.}{%
}{.}\hspace{.4pt}2322823121}}}


\bibitem{ericsson1993protocol}
K.~A. Ericsson and H.~A. Simon.
\newblock {\em Protocol Analysis: Verbal Reports as Data}.
\newblock MIT Press, Cambridge, MA, USA, rev. ed., 1993. \href{https://doi.org/10.7551/mitpress/5657.001.0001}
{doi: {{%
10\hspace{.1pt}\discretionary{.}{%
}{.}\hspace{.4pt}7551\discretionary{/}{%
}{/}mitpress\discretionary{/}{%
}{/}5657\hspace{.1pt}\discretionary{.}{%
}{.}\hspace{.4pt}001\hspace{.1pt}\discretionary{.}{%
}{.}\hspace{.4pt}0001}}}


\bibitem{fu2024data}
Y.~Fu, S.~Guo, J.~Hoffswell, V.~S. Bursztyn, R.~Rossi, and J.~Stasko.
\newblock "{T}he data says otherwise" — towards automated fact-checking and communication of data claims.
\newblock In {\em Proc.\ {ACM} Symp. User Interface Softw. Technol. ({UIST})}, UIST '24,  art. no. 134,  20 pp. {ACM}, New York, Oct. 2024. \href{https://doi.org/10.1145/3654777.3676359}
{doi: {{%
10\hspace{.1pt}\discretionary{.}{%
}{.}\hspace{.4pt}1145\discretionary{/}{%
}{/}3654777\hspace{.1pt}\discretionary{.}{%
}{.}\hspace{.4pt}3676359}}}


\bibitem{goddard2012automation}
K.~Goddard, A.~Roudsari, and J.~C. Wyatt.
\newblock Automation bias: a systematic review of frequency, effect mediators, and mitigators.
\newblock {\em J. Amer. Med. Inform. Assoc.}, 19(1):121--127, Jan. 2012. \href{https://doi.org/10.1136/amiajnl-2011-000089}
{doi: {{%
10\hspace{.1pt}\discretionary{.}{%
}{.}\hspace{.4pt}1136\discretionary{/}{%
}{/}amiajnl\discretionary{%
}{-}{-}2011\discretionary{%
}{-}{-}000089}}}


\bibitem{graves2018understanding}
L.~Graves.
\newblock Understanding the promise and limits of automated fact-checking.
\newblock Factsheet, Reuters Inst. Study Journalism, Univ. Oxford, Oxford, U.K., Feb. 2018. \href{https://doi.org/10.60625/risj-nqnx-bg89}
{doi: {{%
10\hspace{.1pt}\discretionary{.}{%
}{.}\hspace{.4pt}60625\discretionary{/}{%
}{/}risj\discretionary{%
}{-}{-}nqnx\discretionary{%
}{-}{-}bg89}}}


\bibitem{gu2022pasta}
Z.~Gu, J.~Fan, N.~Tang, P.~Nakov, X.~Zhao, and X.~Du.
\newblock {PASTA}: Table-operations aware fact verification via sentence-table cloze pre-training.
\newblock In {\em Proc.\ Conf. Empirical Methods Natural Lang. Process. ({EMNLP})}, pp. 4971--4983. Association for Computational Linguistics, Abu Dhabi, United Arab Emirates, Dec. 2022. \href{https://doi.org/10.18653/v1/2022.emnlp-main.331}
{doi: {{%
10\hspace{.1pt}\discretionary{.}{%
}{.}\hspace{.4pt}18653\discretionary{/}{%
}{/}v1\discretionary{/}{%
}{/}2022\hspace{.1pt}\discretionary{.}{%
}{.}\hspace{.4pt}emnlp\discretionary{%
}{-}{-}main\hspace{.1pt}\discretionary{.}{%
}{.}\hspace{.4pt}331}}}


\bibitem{guo2022survey}
Z.~Guo, M.~Schlichtkrull, and A.~Vlachos.
\newblock A survey on automated fact-checking.
\newblock {\em Trans. Assoc. Comput. Linguistics}, 10:178--206, Feb. 2022. \href{https://doi.org/10.1162/tacl_a_00454}
{doi: {{%
10\hspace{.1pt}\discretionary{.}{%
}{.}\hspace{.4pt}1162\discretionary{/}{%
}{/}tacl\_a\_00454}}}


\bibitem{hassan2017claimbuster}
N.~Hassan, F.~Arslan, C.~Li, and M.~Tremayne.
\newblock Toward automated fact-checking: Detecting check-worthy factual claims by {ClaimBuster}.
\newblock In {\em Proc.\ {ACM} {SIGKDD} Int. Conf. Knowl. Discovery Data Mining ({KDD})}, KDD '17, pp. 1803--1812. {ACM}, New York, Aug. 2017. \href{https://doi.org/10.1145/3097983.3098131}
{doi: {{%
10\hspace{.1pt}\discretionary{.}{%
}{.}\hspace{.4pt}1145\discretionary{/}{%
}{/}3097983\hspace{.1pt}\discretionary{.}{%
}{.}\hspace{.4pt}3098131}}}


\bibitem{horstmann2025trex}
T.~L. Horstmann, B.~Geisenberger, and M.~Alam.
\newblock {T-REX}: Table – refute or entail explainer.
\newblock In {\em Proc.\ Eur. Conf. Mach. Learn. Principles Pract. Knowl. Discovery Databases ({ECML PKDD})}, vol. 16022 of {\em Lecture Notes in Computer Science}, pp. 470--474. Springer, Porto, Portugal, Sep. 2025. \href{https://doi.org/10.1007/978-3-032-06129-4_33}
{doi: {{%
10\hspace{.1pt}\discretionary{.}{%
}{.}\hspace{.4pt}1007\discretionary{/}{%
}{/}978\discretionary{%
}{-}{-}3\discretionary{%
}{-}{-}032\discretionary{%
}{-}{-}06129\discretionary{%
}{-}{-}4\_33}}}


\bibitem{huang2026facilitating}
Y.~Huang and A.~Narechania.
\newblock Facilitating proactive and reactive guidance for decision making on the web: A design probe with {WebSeek}.
\newblock In {\em Proc.\ {ACM} {CHI} Conf. Human Factors Comput. Syst. ({CHI})}, CHI '26,  art. no. 800. {ACM}, New York, Apr. 2026. \href{https://doi.org/10.1145/3772318.3791945}
{doi: {{%
10\hspace{.1pt}\discretionary{.}{%
}{.}\hspace{.4pt}1145\discretionary{/}{%
}{/}3772318\hspace{.1pt}\discretionary{.}{%
}{.}\hspace{.4pt}3791945}}}


\bibitem{hyland2005stance}
K.~Hyland.
\newblock Stance and engagement: A model of interaction in academic discourse.
\newblock {\em Discourse Stud.}, 7(2):173--192, May 2005. \href{https://doi.org/10.1177/1461445605050365}
{doi: {{%
10\hspace{.1pt}\discretionary{.}{%
}{.}\hspace{.4pt}1177\discretionary{/}{%
}{/}1461445605050365}}}


\bibitem{jahanbakhsh2024browser}
F.~Jahanbakhsh and D.~R. Karger.
\newblock A browser extension for in-place signaling and assessment of misinformation.
\newblock In {\em Proc.\ {ACM} {CHI} Conf. Human Factors Comput. Syst. ({CHI})}, CHI '24,  art. no. 946. {ACM}, New York, May 2024. \href{https://doi.org/10.1145/3613904.3642473}
{doi: {{%
10\hspace{.1pt}\discretionary{.}{%
}{.}\hspace{.4pt}1145\discretionary{/}{%
}{/}3613904\hspace{.1pt}\discretionary{.}{%
}{.}\hspace{.4pt}3642473}}}


\bibitem{jonesjang2023ai}
S.~M. Jones-Jang and Y.~J. Park.
\newblock How do people react to {AI} failure? automation bias, algorithmic aversion, and perceived controllability.
\newblock {\em J. Comput.-Mediated Commun.}, 28(1),  art. no. zmac029, Jan. 2023. \href{https://doi.org/10.1093/jcmc/zmac029}
{doi: {{%
10\hspace{.1pt}\discretionary{.}{%
}{.}\hspace{.4pt}1093\discretionary{/}{%
}{/}jcmc\discretionary{/}{%
}{/}zmac029}}}


\bibitem{juneja2022human}
P.~Juneja and T.~Mitra.
\newblock Human and technological infrastructures of fact-checking.
\newblock {\em Proc. ACM Hum.-Comput. Interact.}, 6(CSCW2),  art. no. 418, Nov. 2022. \href{https://doi.org/10.1145/3555143}
{doi: {{%
10\hspace{.1pt}\discretionary{.}{%
}{.}\hspace{.4pt}1145\discretionary{/}{%
}{/}3555143}}}


\bibitem{karagiannis2020scrutinizer}
G.~Karagiannis, M.~Saeed, P.~Papotti, and I.~Trummer.
\newblock Scrutinizer: a mixed-initiative approach to large-scale, data-driven claim verification.
\newblock {\em Proc. VLDB Endowment}, 13(12):2508--2521, Aug. 2020. \href{https://doi.org/10.14778/3407790.3407841}
{doi: {{%
10\hspace{.1pt}\discretionary{.}{%
}{.}\hspace{.4pt}14778\discretionary{/}{%
}{/}3407790\hspace{.1pt}\discretionary{.}{%
}{.}\hspace{.4pt}3407841}}}


\bibitem{choi2024emphasischecker}
D.~H. Kim, S.~Choi, J.~Kim, V.~Setlur, and M.~Agrawala.
\newblock {EmphasisChecker}: A tool for guiding chart and caption emphasis.
\newblock {\em IEEE Trans. Visual. Comput. Graphics}, 30(1):120--130, Jan. 2024. \href{https://doi.org/10.1109/TVCG.2023.3327150}
{doi: {{%
10\hspace{.1pt}\discretionary{.}{%
}{.}\hspace{.4pt}1109\discretionary{/}{%
}{/}TVCG\hspace{.1pt}\discretionary{.}{%
}{.}\hspace{.4pt}2023\hspace{.1pt}\discretionary{.}{%
}{.}\hspace{.4pt}3327150}}}


\bibitem{kim2018texttable}
D.~H. Kim, E.~Hoque, J.~Kim, and M.~Agrawala.
\newblock Facilitating document reading by linking text and tables.
\newblock In {\em Proc.\ {ACM} Symp. User Interface Softw. Technol. ({UIST})}, UIST '18, pp. 423--434. {ACM}, New York, Oct. 2018. \href{https://doi.org/10.1145/3242587.3242617}
{doi: {{%
10\hspace{.1pt}\discretionary{.}{%
}{.}\hspace{.4pt}1145\discretionary{/}{%
}{/}3242587\hspace{.1pt}\discretionary{.}{%
}{.}\hspace{.4pt}3242617}}}


\bibitem{kim2024datadive}
H.~Kim, K.~D. Le, G.~Lim, D.~H. Kim, Y.~J. Hong, and J.~Kim.
\newblock {DataDive}: Supporting readers' contextualization of statistical statements with data exploration.
\newblock In {\em Proc.\ {ACM} Int. Conf. Intelligent User Interfaces ({IUI})}, IUI '24, pp. 623--639. {ACM}, New York, Mar. 2024. \href{https://doi.org/10.1145/3640543.3645155}
{doi: {{%
10\hspace{.1pt}\discretionary{.}{%
}{.}\hspace{.4pt}1145\discretionary{/}{%
}{/}3640543\hspace{.1pt}\discretionary{.}{%
}{.}\hspace{.4pt}3645155}}}


\bibitem{lee2025impact}
H.-P. Lee, A.~Sarkar, L.~Tankelevitch, I.~Drosos, S.~Rintel, R.~Banks et al.
\newblock The impact of generative {AI} on critical thinking: Self-reported reductions in cognitive effort and confidence effects from a survey of knowledge workers.
\newblock In {\em Proc.\ {ACM} {CHI} Conf. Human Factors Comput. Syst. ({CHI})}, CHI '25,  art. no. 1121. {ACM}, New York, Apr. 2025. \href{https://doi.org/10.1145/3706598.3713778}
{doi: {{%
10\hspace{.1pt}\discretionary{.}{%
}{.}\hspace{.4pt}1145\discretionary{/}{%
}{/}3706598\hspace{.1pt}\discretionary{.}{%
}{.}\hspace{.4pt}3713778}}}


\bibitem{lim2023xai}
G.~Lim and S.~T. Perrault.
\newblock {XAI} in automated fact-checking? {T}he benefits are modest and there's no one-explanation-fits-all.
\newblock In {\em Proc.\ Aust. Comput.-Human Interaction Conf. ({OzCHI})}, OzCHI '23, pp. 624--638. {ACM}, New York, May. 2024. \href{https://doi.org/10.1145/3638380.3638388}
{doi: {{%
10\hspace{.1pt}\discretionary{.}{%
}{.}\hspace{.4pt}1145\discretionary{/}{%
}{/}3638380\hspace{.1pt}\discretionary{.}{%
}{.}\hspace{.4pt}3638388}}}


\bibitem{liu2026behavioral}
C.~Liu, Q.~Zhou, X.~Shen, X.~B. Liu, T.~Wu, and X.~A. Chen.
\newblock Behavioral indicators of overreliance during interaction with conversational language models.
\newblock In {\em Proc.\ {ACM} {CHI} Conf. Human Factors Comput. Syst. ({CHI})}, CHI '26,  art. no. 790. {ACM}, New York, Apr. 2026. \href{https://doi.org/10.1145/3772318.3790332}
{doi: {{%
10\hspace{.1pt}\discretionary{.}{%
}{.}\hspace{.4pt}1145\discretionary{/}{%
}{/}3772318\hspace{.1pt}\discretionary{.}{%
}{.}\hspace{.4pt}3790332}}}


\bibitem{lloyd2025beyond}
T.~Lloyd, T.~Nguyen, K.~Levy, and M.~Naaman.
\newblock Beyond community notes: A framework for understanding and building crowdsourced context systems for social media.
\newblock In {\em Proc.\ {ACM} {CHI} Conf. Human Factors Comput. Syst. ({CHI})}, CHI '26,  art. no. 332. {ACM}, New York, Apr. 2026. \href{https://doi.org/10.1145/3772318.3791889}
{doi: {{%
10\hspace{.1pt}\discretionary{.}{%
}{.}\hspace{.4pt}1145\discretionary{/}{%
}{/}3772318\hspace{.1pt}\discretionary{.}{%
}{.}\hspace{.4pt}3791889}}}


\bibitem{metropolitansky2025towards}
D.~Metropolitansky and J.~Larson.
\newblock Towards effective extraction and evaluation of factual claims.
\newblock In {\em Proc.\ 63rd Annu. Meeting Assoc. Comput. Linguistics ({ACL})}, pp. 6996--7045. Association for Computational Linguistics, Vienna, Austria, Jul. 2025. \href{https://doi.org/10.18653/v1/2025.acl-long.348}
{doi: {{%
10\hspace{.1pt}\discretionary{.}{%
}{.}\hspace{.4pt}18653\discretionary{/}{%
}{/}v1\discretionary{/}{%
}{/}2025\hspace{.1pt}\discretionary{.}{%
}{.}\hspace{.4pt}acl\discretionary{%
}{-}{-}long\hspace{.1pt}\discretionary{.}{%
}{.}\hspace{.4pt}348}}}


\bibitem{narechania2024provenancewidgets}
A.~Narechania, K.~Odak, M.~El-Assady, and A.~Endert.
\newblock {ProvenanceWidgets}: A library of {UI} control elements to track and dynamically overlay analytic provenance.
\newblock {\em IEEE Trans. Visual. Comput. Graphics}, 31(1):1235--1245, Jan. 2025. \href{https://doi.org/10.1109/TVCG.2024.3456144}
{doi: {{%
10\hspace{.1pt}\discretionary{.}{%
}{.}\hspace{.4pt}1109\discretionary{/}{%
}{/}TVCG\hspace{.1pt}\discretionary{.}{%
}{.}\hspace{.4pt}2024\hspace{.1pt}\discretionary{.}{%
}{.}\hspace{.4pt}3456144}}}


\bibitem{narechania2021nl4dv}
A.~Narechania, A.~Srinivasan, and J.~Stasko.
\newblock {NL4DV}: A toolkit for generating analytic specifications for data visualization from natural language queries.
\newblock {\em IEEE Trans. Visual. Comput. Graphics}, 27(2):369--379, Feb. 2021. \href{https://doi.org/10.1109/TVCG.2020.3030378}
{doi: {{%
10\hspace{.1pt}\discretionary{.}{%
}{.}\hspace{.4pt}1109\discretionary{/}{%
}{/}TVCG\hspace{.1pt}\discretionary{.}{%
}{.}\hspace{.4pt}2020\hspace{.1pt}\discretionary{.}{%
}{.}\hspace{.4pt}3030378}}}


\bibitem{nguyen2018believe}
A.~T. Nguyen, A.~Kharosekar, S.~Krishnan, S.~Krishnan, E.~Tate, B.~C. Wallace et al.
\newblock Believe it or not: Designing a human-{AI} partnership for mixed-initiative fact-checking.
\newblock In {\em Proc.\ {ACM} Symp. User Interface Softw. Technol. ({UIST})}, pp. 189--199. {ACM}, New York, Oct. 2018. \href{https://doi.org/10.1145/3242587.3242666}
{doi: {{%
10\hspace{.1pt}\discretionary{.}{%
}{.}\hspace{.4pt}1145\discretionary{/}{%
}{/}3242587\hspace{.1pt}\discretionary{.}{%
}{.}\hspace{.4pt}3242666}}}


\bibitem{hanselowski2018neural}
Y.~Nie, H.~Chen, and M.~Bansal.
\newblock Combining fact extraction and verification with neural semantic matching networks.
\newblock {\em Proc. AAAI Conf. Artif. Intell.}, 33(1):6859--6866, Jul. 2019. \href{https://doi.org/10.1609/aaai.v33i01.33016859}
{doi: {{%
10\hspace{.1pt}\discretionary{.}{%
}{.}\hspace{.4pt}1609\discretionary{/}{%
}{/}aaai\hspace{.1pt}\discretionary{.}{%
}{.}\hspace{.4pt}v33i01\hspace{.1pt}\discretionary{.}{%
}{.}\hspace{.4pt}33016859}}}


\bibitem{ons2021coviddeaths}
{Office for National Statistics}.
\newblock Deaths involving {COVID-19} by vaccination status, {England}: deaths occurring between 2 {January} and 2 {July} 2021.
\newblock ONS Statistical Bulletin, 2021.
\newblock [Online]. Available: \url{https://www.ons.gov.uk/peoplepopulationandcommunity/birthsdeathsandmarriages/deaths/articles/deathsinvolvingcovid19byvaccinationstatusengland/deathsoccurringbetween2januaryand2july2021}. {Accessed:} Mar. 31, 2026.

\bibitem{zhang2020adaptive}
K.~Okamura and S.~Yamada.
\newblock Adaptive trust calibration for human-{AI} collaboration.
\newblock {\em PLoS ONE}, 15(2),  art. no. e0229132, Feb. 2020. \href{https://doi.org/10.1371/journal.pone.0229132}
{doi: {{%
10\hspace{.1pt}\discretionary{.}{%
}{.}\hspace{.4pt}1371\discretionary{/}{%
}{/}journal\hspace{.1pt}\discretionary{.}{%
}{.}\hspace{.4pt}pone\hspace{.1pt}\discretionary{.}{%
}{.}\hspace{.4pt}0229132}}}


\bibitem{openai2025gpt41}
{OpenAI}.
\newblock {GPT-4.1}.
\newblock OpenAI API model documentation, 2025.
\newblock [Online]. Available: \url{https://developers.openai.com/api/docs/models/gpt-4.1}. {Accessed:} Jul. 27, 2026.

\bibitem{pesaranghader2026hallucination}
A.~Pesaranghader and E.~Li.
\newblock Hallucination detection and mitigation in large language models, 2026.
\newblock arXiv:2601.09929. [Online]. Available: \url{https://arxiv.org/abs/2601.09929}. \href{https://doi.org/10.48550/arXiv.2601.09929}
{doi: {{%
10\hspace{.1pt}\discretionary{.}{%
}{.}\hspace{.4pt}48550\discretionary{/}{%
}{/}arXiv\hspace{.1pt}\discretionary{.}{%
}{.}\hspace{.4pt}2601\hspace{.1pt}\discretionary{.}{%
}{.}\hspace{.4pt}09929}}}


\bibitem{pirolli2005sensemaking}
P.~Pirolli and S.~Card.
\newblock The sensemaking process and leverage points for analyst technology as identified through cognitive task analysis.
\newblock In {\em Proc.\ Int. Conf. Intelligence Analysis}, vol.~5, pp. 2--4. McLean, May 2005.

\bibitem{pirolli2011introduction}
P.~Pirolli and D.~M. Russell.
\newblock Introduction to this special issue on sensemaking.
\newblock {\em Hum.--Comput. Interact.}, 26(1--2):1--8, Jan. 2011. \href{https://doi.org/10.1080/07370024.2011.556557}
{doi: {{%
10\hspace{.1pt}\discretionary{.}{%
}{.}\hspace{.4pt}1080\discretionary{/}{%
}{/}07370024\hspace{.1pt}\discretionary{.}{%
}{.}\hspace{.4pt}2011\hspace{.1pt}\discretionary{.}{%
}{.}\hspace{.4pt}556557}}}


\bibitem{rubinstein2001executive}
J.~S. Rubinstein, D.~E. Meyer, and J.~E. Evans.
\newblock Executive control of cognitive processes in task switching.
\newblock {\em J. Exp. Psychol.: Hum. Percept. Perform.}, 27(4):763--797, Aug. 2001. \href{https://doi.org/10.1037/0096-1523.27.4.763}
{doi: {{%
10\hspace{.1pt}\discretionary{.}{%
}{.}\hspace{.4pt}1037\discretionary{/}{%
}{/}0096\discretionary{%
}{-}{-}1523\hspace{.1pt}\discretionary{.}{%
}{.}\hspace{.4pt}27\hspace{.1pt}\discretionary{.}{%
}{.}\hspace{.4pt}4\hspace{.1pt}\discretionary{.}{%
}{.}\hspace{.4pt}763}}}


\bibitem{samu2026teammate}
M.~S.~S. Samu, N.~Khan, K.~T. Elahi, T.~B. Rahman, M.~R. Islam, and F.~Sadeque.
\newblock {AI} as teammate or tool? a review of human--{AI} interaction in decision support, 2026.
\newblock arXiv:2602.15865. [Online]. Available: \url{https://arxiv.org/abs/2602.15865}. \href{https://doi.org/10.48550/arXiv.2602.15865}
{doi: {{%
10\hspace{.1pt}\discretionary{.}{%
}{.}\hspace{.4pt}48550\discretionary{/}{%
}{/}arXiv\hspace{.1pt}\discretionary{.}{%
}{.}\hspace{.4pt}2602\hspace{.1pt}\discretionary{.}{%
}{.}\hspace{.4pt}15865}}}


\bibitem{schultz2012truth}
D.~E.~P. Schultz.
\newblock Truth goggles: automatic incorporation of context and primary source for a critical media experience.
\newblock {S.M.} thesis, Program Media Arts Sci., School Archit. Planning, Massachusetts Inst. Technol., Cambridge, MA, USA, 2012.
\newblock [Online]. Available: \url{http://hdl.handle.net/1721.1/76530}.

\bibitem{sultanum2023datatales}
N.~Sultanum and A.~Srinivasan.
\newblock {DataTales}: Investigating the use of large language models for authoring data-driven articles.
\newblock In {\em Proc.\ {IEEE} Visualization Visual Analytics ({VIS})}, pp. 231--235. IEEE, Melbourne, Australia, Oct. 2023. \href{https://doi.org/10.1109/VIS54172.2023.00055}
{doi: {{%
10\hspace{.1pt}\discretionary{.}{%
}{.}\hspace{.4pt}1109\discretionary{/}{%
}{/}VIS54172\hspace{.1pt}\discretionary{.}{%
}{.}\hspace{.4pt}2023\hspace{.1pt}\discretionary{.}{%
}{.}\hspace{.4pt}00055}}}


\bibitem{theologitis2025thucy}
M.~Theologitis and D.~Suciu.
\newblock Thucy: An {LLM}-based multi-agent system for claim verification across relational databases.
\newblock In {\em Proc.\ {AAAI} Workshop {LLM}-Based Multi-Agent Syst. ({LaMAS})}. Singapore, Jan. 2026.
\newblock arXiv:2512.03278. \href{https://doi.org/10.48550/arXiv.2512.03278}
{doi: {{%
10\hspace{.1pt}\discretionary{.}{%
}{.}\hspace{.4pt}48550\discretionary{/}{%
}{/}arXiv\hspace{.1pt}\discretionary{.}{%
}{.}\hspace{.4pt}2512\hspace{.1pt}\discretionary{.}{%
}{.}\hspace{.4pt}03278}}}


\bibitem{thorne2018fever}
J.~Thorne, A.~Vlachos, C.~Christodoulopoulos, and A.~Mittal.
\newblock {FEVER}: A large-scale dataset for fact extraction and {VER}ification.
\newblock In {\em Proc.\ Conf. North Amer. Chapter Assoc. Comput. Linguistics: Human Lang. Technol. ({NAACL-HLT})}, pp. 809--819. Association for Computational Linguistics, New Orleans, Jun. 2018. \href{https://doi.org/10.18653/v1/N18-1074}
{doi: {{%
10\hspace{.1pt}\discretionary{.}{%
}{.}\hspace{.4pt}18653\discretionary{/}{%
}{/}v1\discretionary{/}{%
}{/}N18\discretionary{%
}{-}{-}1074}}}


\bibitem{karagiannis2020identification}
A.~Vlachos and S.~Riedel.
\newblock Identification and verification of simple claims about statistical properties.
\newblock In {\em Proc.\ Conf. Empirical Methods Natural Lang. Process. ({EMNLP})}, pp. 2596--2601. Association for Computational Linguistics, Lisbon, Portugal, Sep. 2015. \href{https://doi.org/10.18653/v1/D15-1312}
{doi: {{%
10\hspace{.1pt}\discretionary{.}{%
}{.}\hspace{.4pt}18653\discretionary{/}{%
}{/}v1\discretionary{/}{%
}{/}D15\discretionary{%
}{-}{-}1312}}}


\bibitem{wadden2020scifact}
D.~Wadden, S.~Lin, K.~Lo, L.~L. Wang, M.~van Zuylen, A.~Cohan et al.
\newblock Fact or fiction: Verifying scientific claims.
\newblock In {\em Proc.\ Conf. Empirical Methods Natural Lang. Process. ({EMNLP})}, pp. 7534--7550. Association for Computational Linguistics, Online, Nov. 2020. \href{https://doi.org/10.18653/v1/2020.emnlp-main.609}
{doi: {{%
10\hspace{.1pt}\discretionary{.}{%
}{.}\hspace{.4pt}18653\discretionary{/}{%
}{/}v1\discretionary{/}{%
}{/}2020\hspace{.1pt}\discretionary{.}{%
}{.}\hspace{.4pt}emnlp\discretionary{%
}{-}{-}main\hspace{.1pt}\discretionary{.}{%
}{.}\hspace{.4pt}609}}}


\bibitem{wang2017liar}
W.~Y. Wang.
\newblock ``liar, liar pants on fire'': A new benchmark dataset for fake news detection.
\newblock In {\em Proc.\ 55th Annu. Meeting Assoc. Comput. Linguistics ({ACL})}, pp. 422--426. Association for Computational Linguistics, Vancouver, Canada, Jul. 2017. \href{https://doi.org/10.18653/v1/P17-2067}
{doi: {{%
10\hspace{.1pt}\discretionary{.}{%
}{.}\hspace{.4pt}18653\discretionary{/}{%
}{/}v1\discretionary{/}{%
}{/}P17\discretionary{%
}{-}{-}2067}}}


\bibitem{wang2024chainoftable}
Z.~Wang, H.~Zhang, C.-L. Li, J.~M. Eisenschlos, V.~Perot, Z.~Wang et al.
\newblock {Chain-of-Table}: Evolving tables in the reasoning chain for table understanding.
\newblock In {\em Proc.\ Int. Conf. Learn. Representations ({ICLR})}. Vienna, Austria, May 2024. \href{https://doi.org/10.48550/arXiv.2401.04398}
{doi: {{%
10\hspace{.1pt}\discretionary{.}{%
}{.}\hspace{.4pt}48550\discretionary{/}{%
}{/}arXiv\hspace{.1pt}\discretionary{.}{%
}{.}\hspace{.4pt}2401\hspace{.1pt}\discretionary{.}{%
}{.}\hspace{.4pt}04398}}}


\bibitem{wongsuphasawat2016voyager}
K.~Wongsuphasawat, D.~Moritz, A.~Anand, J.~Mackinlay, B.~Howe, and J.~Heer.
\newblock Voyager: Exploratory analysis via faceted browsing of visualization recommendations.
\newblock {\em IEEE Trans. Visual. Comput. Graphics}, 22(1):649--658, Jan. 2016. \href{https://doi.org/10.1109/TVCG.2015.2467191}
{doi: {{%
10\hspace{.1pt}\discretionary{.}{%
}{.}\hspace{.4pt}1109\discretionary{/}{%
}{/}TVCG\hspace{.1pt}\discretionary{.}{%
}{.}\hspace{.4pt}2015\hspace{.1pt}\discretionary{.}{%
}{.}\hspace{.4pt}2467191}}}


\bibitem{wongsuphasawat2017voyager2}
K.~Wongsuphasawat, Z.~Qu, D.~Moritz, R.~Chang, F.~Ouk, A.~Anand et al.
\newblock Voyager 2: Augmenting visual analysis with partial view specifications.
\newblock In {\em Proc.\ {ACM} {CHI} Conf. Human Factors Comput. Syst. ({CHI})}, pp. 2648--2659. {ACM}, New York, May 2017. \href{https://doi.org/10.1145/3025453.3025768}
{doi: {{%
10\hspace{.1pt}\discretionary{.}{%
}{.}\hspace{.4pt}1145\discretionary{/}{%
}{/}3025453\hspace{.1pt}\discretionary{.}{%
}{.}\hspace{.4pt}3025768}}}


\bibitem{ye2023dater}
Y.~Ye, B.~Hui, M.~Yang, B.~Li, F.~Huang, and Y.~Li.
\newblock Large language models are versatile decomposers: Decomposing evidence and questions for table-based reasoning.
\newblock In {\em Proc.\ {ACM} Int. Conf. Res. Develop. Inf. Retrieval ({SIGIR})}, pp. 174--184. {ACM}, New York, Jul. 2023. \href{https://doi.org/10.1145/3539618.3591708}
{doi: {{%
10\hspace{.1pt}\discretionary{.}{%
}{.}\hspace{.4pt}1145\discretionary{/}{%
}{/}3539618\hspace{.1pt}\discretionary{.}{%
}{.}\hspace{.4pt}3591708}}}


\bibitem{yuan2023large}
Z.~Yuan, H.~Yuan, C.~Tan, W.~Wang, and S.~Huang.
\newblock How well do large language models perform in arithmetic tasks?, 2023.
\newblock arXiv:2304.02015. [Online]. Available: \url{https://arxiv.org/abs/2304.02015}. \href{https://doi.org/10.48550/arXiv.2304.02015}
{doi: {{%
10\hspace{.1pt}\discretionary{.}{%
}{.}\hspace{.4pt}48550\discretionary{/}{%
}{/}arXiv\hspace{.1pt}\discretionary{.}{%
}{.}\hspace{.4pt}2304\hspace{.1pt}\discretionary{.}{%
}{.}\hspace{.4pt}02015}}}


\bibitem{zamith2019transparency}
R.~Zamith.
\newblock Transparency, interactivity, diversity, and information provenance in everyday data journalism.
\newblock {\em Digit. Journalism}, 7(4):470--489, Jan 2019. \href{https://doi.org/10.1080/21670811.2018.1554409}
{doi: {{%
10\hspace{.1pt}\discretionary{.}{%
}{.}\hspace{.4pt}1080\discretionary{/}{%
}{/}21670811\hspace{.1pt}\discretionary{.}{%
}{.}\hspace{.4pt}2018\hspace{.1pt}\discretionary{.}{%
}{.}\hspace{.4pt}1554409}}}


\end{thebibliography}

\appendix %

\end{document}